\documentstyle[12pt,indent,epsf,eqsection,subeqnarray]{article}

\begin{document}

\title{Lattice Perturbation Theory\break
       for $O(N)$-Symmetric $\sigma$-Models\break
       with General Nearest-Neighbour Action \\[8mm]
       {\large\bf  I. Conventional Perturbation Theory}
}
\author{
  \\
  {\small Sergio Caracciolo}              \\[-0.2cm]
  {\small\it Scuola Normale Superiore and INFN -- Sezione di Pisa}  \\[-0.2cm]
  {\small\it Piazza dei Cavalieri}        \\[-0.2cm]
  {\small\it Pisa 56100, ITALIA}          \\[-0.2cm]
  {\small Internet: {\tt CARACCIO@UX1SNS.SNS.IT}}     \\[-0.2cm]
  {\small Bitnet:   {\tt CARACCIO@IPISNSVA.BITNET}}   \\[-0.2cm]
  {\small Hepnet/Decnet:   {\tt 39198::CARACCIOLO}}   \\[-0.2cm]
  \\[-0.1cm]  \and
  {\small Andrea Pelissetto
   \thanks{
      Address until August 31, 1994:
      Department of Physics, New York University, 4 Washington Place,
      New York, USA; e-mail:
      {\tt PELISSET@MAFALDA.PHYSICS.NYU.EDU}. }}              \\[-0.2cm]
  {\small\it Dipartimento di Fisica and INFN -- Sezione di Pisa}    \\[-0.2cm]
  {\small\it Universit\`a degli Studi di Pisa}        \\[-0.2cm]
  {\small\it Pisa 56100, ITALIA}          \\[-0.2cm]
  {\small Internet: {\tt PELISSET@SUNTHPI1.DIFI.UNIPI.IT}}   \\[-0.2cm]
  {\small Bitnet:   {\tt PELISSET@IPISNSVA.BITNET}}   \\[-0.2cm]
  {\small Hepnet/Decnet:   {\tt 39198::PELISSETTO}}   \\[-0.2cm]
  {\protect\makebox[5in]{\quad}}  
  \\
}
\vspace{0.5cm}

\maketitle
\thispagestyle{empty}   

\vspace{0.2cm}
\clearpage
\begin{abstract}
We compute the $\beta$-function and the anomalous dimension of all
the non-derivative operators of the theory up to three-loops for the
most general nearest-neighbour $O(N)$-invariant action together with some
contributions to the four-loop $\beta$-function. These results are used
to compute the first analytic corrections to various long-distance quantities
as the correlation length and the general spin-$n$ susceptibility.
It is found that these corrections are extremely large for $RP^{N-1}$ models
(especially for small values of $N$),
so that asymptotic scaling can be observed in these models only
at very large values
of $\beta$. We also give the first three terms in the asymptotic expansion of
the vector and tensor energies.
\end{abstract}

\clearpage

\newcommand{\be}{\begin{equation}}
\newcommand{\ee}{\end{equation}}
\newcommand{\<}{\langle}
\renewcommand{\>}{\rangle}

\def\spose#1{\hbox to 0pt{#1\hss}}
\def\ltapprox{\mathrel{\spose{\lower 3pt\hbox{$\mathchar"218$}}
 \raise 2.0pt\hbox{$\mathchar"13C$}}}
\def\gtapprox{\mathrel{\spose{\lower 3pt\hbox{$\mathchar"218$}}
 \raise 2.0pt\hbox{$\mathchar"13E$}}}

\def\bsigma{\mbox{\protect\boldmath $\sigma$}}
\def\bpi{\mbox{\protect\boldmath $\pi$}}
\def\btau{\mbox{\protect\boldmath $\tau$}}
\def\hatp{\hat p}
\def\hatl{\hat l}

\def\msbar{ {\overline{\hbox{\scriptsize MS}}} }
\def\normalmsbar{ {\overline{\hbox{\normalsize MS}}} }

\newcommand{\R}{\hbox{{\rm I}\kern-.2em\hbox{\rm R}}}

\newcommand{\reff}[1]{(\ref{#1})}

\section{Introduction}

The nonlinear $\sigma$-model provides the simplest example
for the realization of a nonabelian global symmetry. It has
found direct applications in condensed-matter physics and its two-dimensional
version has been
extensively studied because it shares with four dimensional gauge
theories the property of being asymptotically free in the
weak-coupling perturbative
expansion~\cite{Polyakov_75,Brezin_76,Bardeen_76}.
This picture predicts a nonperturbative generation of a mass gap which
controls the exponential decay at large distance of the correlation
functions. Thus the two dimensional nonlinear $\sigma$-model has become
an accessible arena for all the attempts to make use of nonperturbative
methods.

The simplest example is the $\sigma$-model where the fields
take values in the sphere $S^{N-1}$
and where the action is invariant under global $O(N)$ transformations.
Besides perturbation theory, it can be studied using different techniques.
It can be solved in the $N=\infty$ limit~\cite{Stanley,DiVecchia}
and $1/N$ corrections can be systematically
computed~\cite{Muller,Flyvbjerg,Campostrini_90ab}.  Moreover an exact
$S$-matrix
can be computed~\cite{Zamolodchikov_79,Polyakov-Wiegmann_83}
and recently , using the thermodynamic Bethe ansatz , the exact mass-gap
of the theory in the limit $\beta\to\infty$ has been obtained
\cite{Hasenfratz-Niedermayer_1,Hasenfratz-Niedermayer_2,Hasenfratz-Niedermayer_3}.
The model can also be studied numerically: by the use of new
collective-mode algorithms which show
little \cite{MGMC_O4} or almost no \cite{Wolff_89a,Wolff_89b,CEPS_swwo4c2}
critical slowing-down,
it has been possible to study the model on very large lattices up to
$\xi \approx 300$. Extensive simulations have been performed
for $N=3$ \cite{Wolff_90,Apostolakis_91},
$N=4$ \cite{Wolff_O4_O8,MGMC_O4}
and $N=8$ \cite{Wolff_O4_O8}.
The results for the correlation length agree with the conventional predictions
--- including the nonperturbative coefficient --- with increasing
accuracy with larger values of $N$, indeed
to within about 20\% for $N=3$, 10\% for $N=4$ and 2\% for $N=8$.

The agreement between theory and Monte Carlo is however not so good if one
considers the so-called variant actions. Consider for instance the Hamiltonian
\be
 \label{eq1.1}
H = - {\beta_V} \sum_{x\mu} (\bsigma_x \cdot \bsigma_{x+\mu}) \, -\,
{\beta_T\over2} \sum_{x\mu} (\bsigma_x \cdot \bsigma_{x+\mu})^2   \;.
\ee
In the formal continuum limit this action reduces to the the usual continuum
$\sigma$-model action with $\beta=\beta_V+\beta_T$.
For $\beta_T=0$, $H$ is the standard $N$-vector Hamiltonian,
while for $\beta_V = 0$
we get the so-called $RP^{N-1}$ theory. A standard one-loop lattice
perturbative computation
\cite{Parisi_80} gives the asymptotic (as $\beta\to\infty$)
ratio between the correlation lengths of these two models.
However, already the first simulations showed a large discrepancy between the
Monte Carlo and the theoretical prediction~\cite{Sinclair_82}.
In a recent study~\cite{CEPS_LAT92} of the $RP^2$ theory on lattices of size
$L\leq 512$, at $\beta_T =5.58$ we measure $\xi_T \approx 200$,
while theory predicts $\xi_T \approx 10^9$.

Various more-or-less-radical theoretical ideas have been proposed
to account for this discrepancy.
The theory with $\beta_V = 0$ enjoys a $Z_2$ gauge invariance:
the Hamiltonian is invariant under local inversions of the spins, so that
the fields really belong to the manifold $RP^{N-1}= S^{N-1}/Z_2$. It has been
argued that this discrete gauge symmetry,
which should be irrelevant in the perturbative limit,
changes the non-perturbative behaviour and gives rise to a phase transition
with infinite correlation length at a finite value of $\beta_T$
\cite{Solomon_81,Caselle-Gliozzi-Megna,Fukugita_82,Kunz_91}.
This scenario is of course incompatible with the conventional
asymptotic-freedom (or more precisely infrared-slavery) predictions.
Indeed, some plausible arguments have been put forward in
order to show that the $\sigma$-model --- even the standard one
with $\beta_T = 0$ --- is not asymptotically free, but rather has a
low-temperature phase with zero mass gap and algebraically decaying correlation
functions~\cite{Patrascioiu_92b,Seiler}.

In this paper we try to understand whether these discrepancies
between theory and ``experiment'' can be understood
within {\em conventional}\/ perturbation theory.
We do this by extending to general nearest-neighbour actions
--- including \reff{eq1.1} ---
the previous two-loop and three-loop lattice computations.
{}From the three-loop corrections we can obtain a rough estimate of
how large $\beta$ has to be in order for the perturbative predictions
to be reliable.

After introducing in Section~2  our notations,
we give in Section~3
the expectation values for the vector and tensor energies up to three
loops, thus extending previous results of L\"uscher~\cite{Luscher_unpub}.
This calculation can be of help, according to the ideas first presented
in~\cite{Parisi_Madison},  if the numerical data are analyzed in terms,
instead of the bare coupling, of an effective coupling which is a
function of the measured averaged value of a local observable.
Our main results are presented in Section~4 where we compute
the asymptotic expansion for large $\beta$ of the correlation length and of
the general spin-$n$ susceptibility:
in subsection 4.1 we give an analytic
expression for the three-loop $\beta$-function on the lattice with
general nearest-neighbour action
generalizing the computation reported
in~\cite{Falcioni-Treves} for the standard $N$-vector action
(in addition we are able to compute exactly one of the
independent integrals they used to parametrize their result).
Moreover we compute some of the contributions
to the four-loop $\beta$-function.: these terms are probably the largest ones
for the $RP^{N-1}$-model defined by \reff{eq1.1}.
Using  these results we can compute the first analytic correction to the
correlation length and some of the contributions to the second analytic
correction: it turns out that these terms are extremely large
for the $RP^{N-1}$ at correlation lengths of order 10-100 which is
the range studied in Monte Carlo simulations. In particular for
the theory defined by \reff{eq1.1} with $\beta_V=0$ and $N=3$ we find that the
perturbative formula is unreliable for $\beta_T\ltapprox 15$ which
is very far from the region studied in Monte Carlo simulations
($\beta_T\ltapprox 5.6$).
Therefore this could provide a theoretical
explanation of the reason why asymptotic scaling is not seen in
Monte Carlo simulations.
In subsection 4.2 we extend the discussion to the susceptibilities.
We compute the anomalous dimension of the general spin-$n$
susceptibility and give the first analytic corrections
to its asymptotic behaviour.
Then in subsection 4.3 we discuss the possibility of parametrizing
long-distance quantities in terms
of local observables~\cite{Parisi_Madison}, showing that by expressing them
in terms of the vector energy or of the square root
of the tensor energy the corrections are significantly smaller
at {\em all} loops.
With this parametrization the discrepancies are expected to be smaller.
A detailed comparison with high-precision Monte Carlo data will be presented in
\cite{CEPS_93}.


In Appendix  A we present some technical details for the computation reported
in
Section 2 while
in Appendix \ref{secB}
we summarize some results of the
$1/N$ expansion~\cite{Magnoli-Ravanini} in two dimensions of the model
\reff{eq1.1}
which provide a useful check for a part of our results.

\section{Notations}   \label{sec2}

\subsection{Definition of the Model}  \label{sec2.1}

In this Section we define our general model.
We will  be
considering the most general  $O(N)$-invariant nearest-neighbour interaction
for unit-length spins $\bsigma_x \in \R^N$, i.e. a lattice
Hamiltonian of the form
\begin{eqnarray}
H^{latt} & = & - \sum_{x\mu} {\cal V} \left(
\bsigma_x\cdot\bsigma_{x+\mu}\right)
  \label{eq2.1}  \\
& = & - \sum_{x\mu} {\cal V} \left( 1 - (\Delta_\mu\bsigma)^2/2\right)   \;,
\end{eqnarray}
where ${\cal V} \colon\, [-1,1] \to \R$ is a generic function and
\be
\Delta_\mu \bsigma_x = \bsigma_{x+\mu} - \bsigma_x
\ee
where $\mu=1,2,\ldots,d$ are the positive directions on a hypercubic lattice
with lattice spacing $a=1$.

\noindent The partition function is given by
\be
Z \, =\, \int  e^{-\beta H^{latt}}  \prod_x
\hbox{\rm d}^N\bsigma_x \, \delta(\bsigma_x^2 -1) \;.
\ee
In the formal continuum limit the Hamiltonian reduces to
\be
-{\rm const} \, + {1\over2}\; {\cal V'}(1)\, \int \hbox{d}^d x\,
\partial_\mu \bsigma \cdot \partial_\mu \bsigma \; .
\ee
In order to have the usual normalizations we will require ${\cal V'}(1) = 1$.

An important requirement on the function ${\cal V}(x)$ is connected with the
nature of the ground state. Indeed we will require the theory to be
ferromagnetic, i.e. we want the configurations with all spins aligned in a
fixed
direction to be absolute minima of the Hamiltonian. This means that
${\cal V}(x)$ must have the property
\be
\max_{x\in [-1,1]}\, {\cal V}(x)\, = {\cal V}(1)  \;. \label{minimum-on-V}
\ee
In principle there could be other points $\bar x$ such that
${\cal V}(\bar x)\, = {\cal V}(1)$.
Let us firstly suppose
that $x=1$ is the {\em unique}\/
absolute maximum of ${\cal V}(x)$ for $x\in [-1,1]$.
Then, for large $\beta$, each nearest-neighbour product
$\bsigma_x \cdot\bsigma_{x+\mu}$ is (with high probability)
of order $1-O(1/\beta)$ and thus we can expand ${\cal V}(x)$ around
$x=1$. We will need ${\cal V}(x)$ through order $(x-1)^3$,
as terms of order $(x-1)^n$ begin to contribute at the $n$-loop level and we
will stop our computation at three loops.
Thus,
\be
{\cal V}(x) \,=\, \hbox{\rm const } + (x-1) + {r\over2}\, (x-1)^2 +
{s\over3!}\,
  (x-1)^3 + \, O((x-1)^4)  \;.
\ee
It follows that
\be
H^{latt}\, =\, {1\over2}  \sum_{x\mu} (\Delta_\mu\bsigma)^2 - {r\over8}
\sum_{x\mu} (\Delta_\mu\bsigma\cdot\Delta_\mu\bsigma)^2
+ {s\over48} \sum_{x\mu} (\Delta_\mu\bsigma\cdot\Delta_\mu\bsigma)^3
\,+\,  O\!\left( (\Delta_\mu\bsigma)^8 \right)   \;.
\ee
Note that for the mixed isovector-isotensor action \reff{eq1.1}
we have $\beta = \beta_V + \beta_T$,
$r = \beta_T/(\beta_V + \beta_T)$ and $s=0$.

Let us now discuss the case in which there are other points $x=\bar x$ such
that
${\cal V}(\bar x)= {\cal V}(1)$. In this case one must consider the
contributions
of all the degenerate minima of the Hamiltonian in order to get the correct
answer. There is, however, a special situation in which this
summation can be avoided. It is the case of the $RP^{N-1}$ models, where
$x=\pm1$ are the only maxima of ${\cal V}(x)$ and where ${\cal V}(x)$ is an
even function of $x$. Such a Hamiltonian has an additional $Z_2$ local
gauge invariance, as the transformations
\be
\bsigma_i \to \, s_i \, \bsigma_i
\ee
with $s_i = \pm 1$ leave it invariant. Thus one can limit oneself
to consider $Z_2$-invariant correlations, and for these functions all the
(infinitely many)
degenerate minima give exactly the same contribution. Consequently
by considering only fluctuations around the configurations with
$\bsigma_x\cdot\bsigma_{x+\mu} \approx 1$ one gets the correct answer.


In order to proceed further we must consider all configurations such that
$\bsigma_x\cdot\bsigma_{x+\mu} = 1$ for all $x,\mu$.
It is obvious that there is an infinite
number of them: indeed any configuration $\bsigma = {\bf v}$ for a fixed
vector ${\bf v}$ is a minimum of the Hamiltonian. In $d>2$ this is not a
serious problem as at large $\beta$ the pure phases of the model exhibit
spontaneous symmetry breaking~\cite{FSS},
thus guaranteeing that if all spins point
in a given direction for $\beta=\infty$, the same holds true at finite
$\beta$ with corrections of order $1/\beta$. The situation is different
in $d \le 2$ as the Mermin-Wagner theorem forbids the spontaneous
breaking of a continuous global symmetry. Technically
this is reflected in the fact that the perturbative expansion is ill-behaved
and plagued by infrared divergences. We do not want to discuss
this problem here (but see Ref.~\cite{CPS_conceptual})
and we will adopt the common technique of adding a magnetic field $h$ to
the Hamiltonian which provides an infrared regulator at the intermediate
stages of the computation. The general results by David~\cite{David} guarantee
that
correlations of $O(N)$-invariant quantities have a finite limit,
order-by-order in perturbation theory, for $h\to 0$.
If the magnetic field points in the first  direction we have thus the
Hamiltonian
\be
H^{latt} \, = \, - \sum_{x\mu} {\cal V} \left(
\bsigma_x\cdot\bsigma_{x+\mu}\right)
- h\, \sum_x \bsigma^1_x   \;.
  \label{eq2.10}
\ee
The introduction of $h$ removes the degeneracy of the ground state
and thus for large $\beta$ we have $\bsigma \approx (1, {\bf 0})$.
In order to perform the perturbative expansion we then set
$\bsigma_x = (\sqrt{1 - \bpi_x^2}, \bpi_x)$
where $\bpi_x \in \R^{N-1}$ and $|\bpi_x| \le 1$.
We must now take into account the jacobian
due to the change of variables in the integration measure:
\begin{eqnarray}
\prod_x \delta\left( \bsigma_x^2 -1 \right) \,d^N\bsigma_x & = &
\prod_x {1\over 2 \sqrt{1 - \bpi_x^2}} \,d^{N-1}\bpi_x \nonumber \\
& = & \left( \prod_x {d^{N-1}\bpi_x \over 2}\right) \exp \left[ -  \sum_x
{1\over 2}
\log \left( 1 - \bpi_x^2\right) \right]
\end{eqnarray}

The exponential term can be included as an effective contribution to the
Hamiltonian,
and an expansion for $\bpi_x^2  \ll 1$ produces
\begin{eqnarray}
  \beta H^{eff}(\bpi)   & \equiv &
  \beta H^{latt}(\bsigma) \,+\,
          {1\over 2} \sum_x \log \left( 1 - \bpi_x^2\right)     \nonumber
\\
   & = &
     {\beta \over 2} \sum_x \left[\sum_\mu \left( \Delta_\mu \bpi_x \right)^2
                                   + h \bpi_x^2 \right]
  \,-\, {1\over 2 } \sum_x \bpi_x^2
  \,+\, {\beta h\over 8} \sum_x \left(\bpi_x^2 \right)^2
                                                                   \nonumber \\
&  & \qquad  \,+\, {\beta\over 8} \sum_{x,\mu}
                        \left( \Delta_\mu \bpi_x^2 \right)^2
             \,-\, {\beta r\over 8} \sum_{x,\mu}
                    \left( \Delta_\mu \bpi_x \cdot \Delta_\mu \bpi_x\right)^2
\nonumber \\
& & \qquad - {1\over4} \sum_x (\bpi^2_x)^2 + {\beta\over16} \sum_{x,\mu}
\left( \Delta_\mu \bpi^2_x \right)^2 (\bpi^2_x + \bpi^2_{x+\mu}) \nonumber \\
& & \qquad - {\beta r\over 16} \sum_{x,\mu}
\left( \Delta_\mu \bpi_x \cdot \Delta_\mu \bpi_x\right)
\left( \Delta_\mu \bpi^2_x \right)^2 \\
& & \qquad + {\beta s \over 48} \sum_{x,\mu}
\left( \Delta_\mu \bpi_x \cdot \Delta_\mu \bpi_x\right)^3
+ {\beta h\over 16} \sum_x (\bpi^2_x)^3 +\, O(\bpi^8) \nonumber
\end{eqnarray}

\subsection{Some Lattice Integrals}   \label{sec2.2}

Let us now introduce a few notations which we will use in the following
paragraphs.
The one-loop
perturbative results will be written in terms of the lattice integral
\be
   I(h) \;=\; \int\limits_{[-\pi,\pi]^d} \! {{\rm d}^d p\over (2 \pi)^d} \;
               {1\over \hatp^2 + h}   \;,
 \label{eq2.13}
\ee
where $\hatp^2$ denotes the useful shorthand
\be
\hatp^2  \;=\; \sum_\mu \hatp^2_\mu  \;=\;
               \sum_\mu \left( 2 \sin {p_\mu\over 2} \right)^2    \;.
\ee
For $d>2$ $I(h)$ is finite when $h\to 0$,
while for $d \le 2$ it diverges.  More precisely,
for $d=2$ we have
\be
I(h) \;=\;
   {2 \over \pi(4+h)}  K\!\left( {4 \over 4+h} \right)
   \;=\;
           -{1\over 4 \pi} \log {h\over 32}  \, + O(h\log h) \; ,
\ee
where $K$ is a complete elliptic integral of the first kind,
while for $d=1$
\be
I(h) \; =\; {1\over\sqrt{4h+h^2}} \;=\; {1\over2\sqrt{h}} + \ O(\sqrt{h})\; .
\ee
Let us also introduce
\be
   I_2(h) \;=\; \int\limits_{[-\pi,\pi]^d} \! {{\rm d}^d p\over (2 \pi)^d} \;
               {1\over (\hatp^2 + h)^2 }   \; = - \, {d I(h) \over
                d h} \; .
\ee

At the two-loop level we shall need the integrals
\begin{eqnarray}
G_1 & = & -{1\over4}\int \limits_{[-\pi,\pi]^d}
{{\rm d}^d p \over (2\pi)^d} {{\rm d}^d q \over (2\pi)^d}
\left[\sum_\mu \widehat{(p+q)^4_\mu}\right]\, {\widehat{(p+q)^2} - \hat{p}^2 -
\hat{q}^2
\over \hat{p}^2 \hat{q}^2 [\widehat{(p+q)^2}]^2 } \\
R & = & \lim_{h\to 0} \ h\, \int \limits_{[-\pi,\pi]^d}
{{\rm d}^d p \over (2\pi)^d} {{\rm d}^d q \over (2\pi)^d}
{1 \over (\hat{p}^2 +h ) (\hat{q}^2 + h)(\widehat{(p+q)^2}+h) }
\end{eqnarray}
The integral $G_1$ already appears in the work by Falcioni and
Treves~\cite{Falcioni-Treves} and we
follow their notation. In two dimensions
\be G_1 \approx 0.0461636\; . \ee
The integral $R$,
which vanishes when $d>2$, appears only in the intermediate stages of the
computation but cancels in any of the results. In two dimensions we have
\be
R \, =\, {1\over 24\pi^2}\, \psi'(1/3) - {1\over 36} \, \approx\,
0.0148430
\ee
where $\psi(z) = d\log\Gamma(z)/dz$.

In the three-loop computation we will express all results in terms of two
integrals:
\be
J = \int\limits_{[-\pi,\pi]^d} {{\rm d}^d p \over (2\pi)^d}
{{\rm d}^d q \over (2\pi)^d}
{{\rm d}^d r \over (2\pi)^d} {{\rm d}^d s \over (2\pi)^d}
(2\pi)^d \delta(p+q+r+s) {(\sum_\mu  \hat{p}_\mu \hat{q}_\mu \hat{r}_\mu
\hat{s}_\mu )^2\over  \hat{p}^2 \hat{q}^2 \hat{r}^2 \hat{s}^2}
\ee
and
\begin{eqnarray}
\lefteqn{K = \int \limits_{[-\pi,\pi]^d}{{\rm d}^d p \over (2\pi)^d}
{{\rm d}^d q \over (2\pi)^d}
{{\rm d}^d r \over (2\pi)^d} {{\rm d}^d s \over (2\pi)^d}
(2\pi)^d \delta(p+q+r+s)    }  \hspace{5cm}\nonumber    \\
& & { [\widehat{(p+q)}^2 - \hat{p}^2 - \hat{q}^2]
[\widehat{(r+s)}^2 - \hat{r}^2 - \hat{s}^2] \over \hat{p}^2 \hat{q}^2
\hat{r}^2 \hat{s}^2}
\end{eqnarray}
Numerically in two dimensions
\begin{eqnarray}
J & \approx & 0.1366198 \\
K & \approx & 0.095888
\end{eqnarray}

In our partial computation of the four-loop $\beta$-function we will also
use the following three-loop two-dimensional integrals:
\begin{eqnarray}
L_1 &=& \int \limits_{[-\pi,\pi]^2}{{\rm d}^2 p \over (2\pi)^2}
{{\rm d}^2 q \over (2\pi)^2}
{{\rm d}^2 r \over (2\pi)^2} {{\rm d}^2 s \over (2\pi)^2}
(2\pi)^2 \delta(p+q+r+s) {1\over (\hat{p}^2)^2 \hat{q}^2
\hat{r}^2 \hat{s}^2 \widehat{(r+s)}^2} \nonumber \\
& & \qquad \sum_{\mu\nu} \widehat{p}_\mu  \widehat{s}_\mu \widehat{r}_\mu
\widehat{p}_\nu  \widehat{s}_\nu \widehat{r}_\nu ( \widehat{q}_\mu
\widehat{q}_\nu \widehat{(r+s)}^2 - \widehat{(r+s)}_\mu  \widehat{(r+s)}_\nu
\hat{q}^2) \\
L_2 &=& \int \limits_{[-\pi,\pi]^2}{{\rm d}^2 p \over (2\pi)^2}
{{\rm d}^2 q \over (2\pi)^2}
{{\rm d}^2 r \over (2\pi)^2} {{\rm d}^2 s \over (2\pi)^2}
(2\pi)^2 \delta(p+q+r+s) {\sum_\mu \hat{p}_\mu \hat{q}_\mu \hat{r}_\mu
\hat{s}_\mu \widehat{(r+s)}^2_\mu\over \hat{p}^2 \hat{q}^2
\hat{r}^2 \hat{s}^2 } \\
L_3 &=& \int \limits_{[-\pi,\pi]^2}{{\rm d}^2 p \over (2\pi)^2}
{{\rm d}^2 q \over (2\pi)^2}
{{\rm d}^2 r \over (2\pi)^2} {{\rm d}^2 s \over (2\pi)^2}
(2\pi)^2 \delta(p+q+r+s) {1\over \hat{p}^2 \hat{q}^2
\hat{r}^2 \hat{s}^2 \widehat{(p+q)}^2} \nonumber \\
& & \qquad \sum_{\mu\nu} \widehat{p}_\mu \widehat{q}_\mu
\widehat{s}_\mu \widehat{r}_\mu
\widehat{p}_\nu  \widehat{q}_\nu \widehat{(p+q)}^2_\nu
\cos\left({p+q\over2}\right)_\nu
\end{eqnarray}
Numerically we get
\begin{eqnarray}
L_1 &\approx& 0.0029333\\
L_2 &\approx& 0.0102157\\
L_3 &\approx& - 0.0167292
\end{eqnarray}

\section{Local Quantities}   \label{sec3}

We wish now to compute the average values of the vector and tensor energy per
link. The expansion in terms of the field $\bpi$ is given by:
\begin{eqnarray}
   E_V  & \equiv &  \< \bsigma_x\cdot\bsigma_{x+\mu} \>           \nonumber \\
        & = &
   1  \,-\, {1\over 2} \< ( \Delta_\mu \bpi_x )^2 \>
      \,-\, {1\over 8}  \< ( \Delta_\mu \bpi_x^2 )^2 \>           \nonumber \\
      & & \quad - {1 \over 16} \< ( \Delta_\mu \bpi_x^2 )^2 (\bpi_x^2 +
                         \bpi^2_{x+\mu})\> \,+\, O\left( \< \bpi^8 \> \right)
                                     \\[4mm]
   E_T  & \equiv &  \< \left(\bsigma_x\cdot\bsigma_{x+\mu} \right)^2\>
                                                                  \nonumber \\
        & = &
   1  \,-\,  \< \left( \Delta_\mu \bpi_x \right)^2 \>
      \,-\, {1\over 4}  \< ( \Delta_\mu \bpi_x^2 )^2 \>
      \,+\, {1\over 4} \< ( \Delta_\mu \bpi_x \cdot \Delta_\mu \bpi_x )^2 \> \\
      & & \quad -  {1 \over 8} \< ( \Delta_\mu \bpi_x^2 )^2 (\bpi_x^2 +
                         \bpi^2_{x+\mu}) \>
      \,+\, {1\over8} \< ( \Delta_\mu \bpi_x )^2 \, ( \Delta_\mu \bpi_x^2
)^2\>\,
       +\, O\left( \< \bpi^8 \> \right)
                                                                  \nonumber
\end{eqnarray}
In Figure~\ref{fig1} we report the Feynman graphs which contribute to the
energies.
\begin{figure}
\vspace*{0cm} \hspace*{-0cm}
\begin{center}
\epsfxsize = 0.8\textwidth
\leavevmode\epsffile{3loops1.ps}  
\end{center}
\vspace*{0cm}
\caption{
  Feynman graphs relevant to the evaluation of the energies respectively to
  order (a) $1/\beta$ (b) $1/\beta^2$ (c) $1/\beta^3$.
  The dot indicates
  the operator insertion while the cross represents the vertices coming
  from the measure term in $H^{eff}$.}
\label{fig1}
\end{figure}
Through second order in $1/\beta$, we get  \begin{eqnarray}
\< ( \Delta_\mu \bpi_x )^2 \>
    & = &   {N-1\over \beta d} (1 -h I(h)) \\
    && + {N-1\over 2\beta^2d^2}\, \left( 1 + r(N+1)\right)(1 - hI(h))
         (1 - 2hI(h) + h^2 I_2(h)) \nonumber \\
    && + {N-1\over \beta^2d}\, \left[-I(h) - {N-5\over2} h I^2(h) +
         {N-3\over2} h^2 I(h) I_2(h) \right] \nonumber \\[3mm]
\< ( \Delta_\mu \bpi_x^2 )^2 \>
    & = & {4(N-1)\over \beta^2d}\left[ I(h) - hI^2(h) - {1\over 4d}(1 -
hI(h))^2
          \right] \\[3mm]
\< ( \Delta_\mu \bpi_x \cdot \Delta_\mu \bpi_x )^2 \>
    & = &   {N^2-1\over  \beta^2 d^2} ( 1 - hI(h))^2
\end{eqnarray}
Let us firstly suppose $d \geq 2$. In this case, in the limit $h\to 0$, we get
\begin{eqnarray}
 E_V   & = &  1  \,-\, {N-1\over 2 \beta d}
                 \,-\, {N-1\over 8 \beta^2 d^2}
                 \,-\, r {N^2-1\over 4 \beta^2 d^2}
                 \,+\, O(1/\beta^3)    \label{eq2.40}                \\[2mm]
 \label{energy_V}
 E_T   & = &  1  \,-\, {N-1\over  \beta d}
                 \,+\, {N(N-1)\over  4 \beta^2 d^2}
                 \,-\, r {N^2-1\over 2 \beta^2 d^2}
                 \,+\, O(1/\beta^3)    \label{eq2.41}
 \label{energy_T}
\end{eqnarray}
Notice the cancellation of the infrared divergent terms, which
is guaranteed to all orders by a general theorem of David \cite{David}.

The formula \reff{energy_T} produces for the $RP^{N-1}$ models the result
\be
E_T  \;=\;  1 \,-\, {N-1\over  \beta d} \,-\, {(N+2) (N-1)\over  4 \beta^2 d^2}
              \,+\, O(1/\beta^3)   \;.
\ee
For $d=2$, $N=3$ this reduces to
\be
E_T  \;=\; 1 \,-\, {1\over \beta} \,-\, {5\over 8 \beta^2}
              \,+\, O(1/\beta^3)   \;,
\ee
in agreement with the result in~\cite{Sinclair_82}.

Let us mention that
these formulae for $E_V$ and $E_T$ are valid only for models in which
$x=1$ is the {\em unique}\/ absolute
maximum of ${\cal V}(x)$. For $RP^{N-1}$ models,
only the result for $E_T$ is correct;
indeed, as the vector energy $E_V$ changes sign
under the gauge transformations,
Elitzur's theorem guarantees that $E_V = 0$ identically.

Let us now consider the case $d=1$. In this case also the terms proportional to
$h I^2(h)$ and to $h^2 I(h) I_2(h)$ contribute in the limit $h\to 0$. We get
\begin{eqnarray}
 E_V   & = &  1  \,-\, {N-1\over 2 \beta }
                 \,+\, {(N-1)(N-7)\over 32 \beta^2}
                 \,-\, r {N^2-1\over 4 \beta^2}
                 \,+\, O(1/\beta^3)       \label{eq2.44}             \\[2mm]
 E_T   & = &  1  \,-\, {N-1\over  \beta}
                 \,+\, {(5N-3)(N-1)\over 16 \beta^2}
                 \,-\, r {N^2-1\over 2 \beta^2}
                 \,+\, O(1/\beta^3)       \label{eq2.45}
\end{eqnarray}
In $d=1$ we can compare with the exact solution. Indeed in this case
the partition function of a system of $L+1$ spins with free boundary conditions
at zero magnetic field is simply
\be
Z =\, W^L\, {\pi^{N/2}\over \Gamma(N/2)}
\ee
where $W$ is the following one-link integral
\be
W\, =\, \int {\rm d}^N\bsigma \, \delta(\bsigma^2 - 1) \exp\left[\beta
{\cal V} (\sigma^1) \right]
\ee
In the limit $\beta\to\infty$,
we get the following asymptotic expansion
\be
W=\, {1\over2} \left({2\pi\over\beta}\right)^{(N-1)/2}\, e^{\beta{\cal V}(1)}
\,
\left[ 1 +{N-1\over 2\beta} - (1-r){N^2-1\over8\beta}  +\,
O(1/\beta^2) \right]
\ee
 From this expression it is easy to get the asymptotic expansion of $E_V$ and
$E_T$:
\begin{eqnarray}
 E_V   & = &  1  \,-\, {N-1\over 2 \beta }
                 \,+\, {(N-1)(N-3)\over 8 \beta^2}
                 \,-\, r {N^2-1\over 4 \beta^2}
                 \,+\, O(1/\beta^3)                                   \\[2mm]
 E_T   & = &  1  \,-\, {N-1\over  \beta}
                 \,+\, {(N-1)^2\over  2 \beta^2}
                 \,-\, r {N^2-1\over 2 \beta^2}
                 \,+\, O(1/\beta^3)
\end{eqnarray}
Notice that these expressions do {\em not}\/ coincide
with the expressions \reff{eq2.44}/\reff{eq2.45} obtained using
standard perturbation theory. This is a reflection of the fact that the
limits $h\to0$ and $\beta\to\infty$ do not
commute~\cite{Rossi-Brihaye,Hasenfratz_1d,CPS_conceptual}.

Let us now compute the $O(1/\beta^{3})$-corrections to $E_V$ and $E_T$
keeping only those terms which are non-zero for $h\to 0$ and $d\geq2$.
Here we indicate with $\<A\>_{3l}$ the coefficient of
$1/\beta^3$ in $\<A\>$. We obtain
\begin{eqnarray}
\< ( \Delta_\mu \bpi_x )^2 \>_{3l}
    & = &   -{(N-1)^2\over 2 d} \left[ {1\over 2d} I(h) - K\right] \nonumber \\
    & & \quad -
    {N-1\over d}\left[ - I(h)^2 + {1\over d} I(h) - {1\over 3d^2} + {K\over2}
    - {J\over6}\right] \nonumber \\
    & & \quad - {N^2-1\over d^2} r\left[ I(h) - {7\over 24 d} - {d\over3}J
\right]  \\
    & & \quad + {N^2-1\over 2 d^3} r^2 (N+1+d^2 J)
- {s\over 8d^3} (N^2-1)(N+3) \nonumber\\
    & & \quad + {N-1\over 4}\left[ (N-3)I(h) + {1\over2}\right] hI_2(h)
\nonumber \\
    & & \quad + {N^2-1\over8} r hI_2(h) \nonumber \\
\< ( \Delta_\mu \bpi^2_x )^2 \>_{3l}
    & = & -{(N-1)^2\over d}\left[4 I(h)^2- {2\over d} I(h) + K\right]\nonumber
\\
    & & \quad -
   {4\over d}(N-1)\left[3 I(h)^2 - {3\over2d}I(h) + {1\over 6d^2} - {K\over4}
    + {J\over 12}\right] \nonumber \\
    & & \quad + {4\over d^2}(N^2-1) r\left[I(h) + {1\over 6 d} - {d\over 12} J
    \right] \\
    & & \quad - (N-1) \left[ (N-3)I(h) + {1\over 2}\right] hI_2(h) \nonumber \\
    & & \quad - {N^2-1\over 2} r hI_2(h) \nonumber \\
\< ( \Delta_\mu \bpi^2_x )^2 (\bpi^2_x + \bpi^2_{x+\mu})\>_{3l}
    & = & {8\over d}(N^2-1)\left[ I(h)^2 - {1\over 4d} I(h)\right]  \\
\< [( \Delta_\mu \bpi_x )^2]^2 \>_{3l}
    & = & {N^2-1\over d}\left[-{2\over d}I(h) + {2\over 3 d^2} +
{J\over3}\right]
    \nonumber \\
    & & \quad + {N^2-1\over d^3} r (N+1+d^2 J) \\
\< ( \Delta_\mu \bpi_x )^2 ( \Delta_\mu \bpi^2_x )^2\>_{3l}
    & = & {4\over d^2} (N^2-1) \left( I(h) - {1\over 4d}\right)
\end{eqnarray}
We thus get for the $1/\beta^3$ term of $E_V$ and $E_T$:
\begin{eqnarray}
(E_V)_{3l} & = & -{(N-1)^2\over 8d} K + {N-1\over 2d} \left({K\over4} -
{J\over12} - {1\over 6d^2}\right) \nonumber \\
&& - {N^2-1\over 16d^3} r (1 + 2 d^2 J) - {N^2-1\over 4d^3} r^2 (N+1 + d^2 J)
\nonumber \\
&& + {s\over 16d^3}\, (N^2-1)(N+3) \label{eq2.56} \\
(E_T)_{3l} & = & {(N-1)^2\over 4d} \left( - K +{1\over 6d^2} + {J\over3}\right)
+ {N-1\over 4d}\left( K - {1\over 3d^2} +{J\over3}\right) \nonumber \\
&& + {N^2-1\over 8d^3} (2N+1) r - {N^2-1\over 2d^3} r^2 (N+1 + d^2 J)
\nonumber \\
&& + {s\over 8d^3}\, (N^2-1)(N+3) \label{eq2.57}
\end{eqnarray}
Notice that the final result does not contain terms proportional to $I(h)$,
which would diverge in two dimensions, nor terms proportional to $hI_2(h)$,
which
is non-vanishing for $h\to 0$ only in two dimensions.

Several checks can be made on this result:
\begin{itemize}
\item[1)] For $N=2$ we have
\be
E_T \left(\beta,r=1,s=0\right) \, =\, {1\over2} + {1\over2}
E_V \!\left({\beta\over4},r=0,s=0\right)   \;,
\ee
an identity which is easily derived by noticing that $RP^1$ and $S^1$ define
the same theory.
\item[2)]  In the limit $N\to\infty$ with $\beta/N$ fixed
and $s=0$ (and $d=2$), we can compare to the results in Appendix \ref{secB}
for the large-$N$ limit of the
mixed isovector-isotensor action \reff{eq1.1}.
Indeed, \reff{eq2.40}/\reff{eq2.56} and \reff{eq2.41}/\reff{eq2.57}
agree with \reff{eqC.15} and \reff{eqC.16}, respectively.
Of course, this checks only the terms with the highest power of $N$
at each order, i.e.\ $N^\ell/\beta^\ell$.
For the standard action $(r=0)$ we can also compare
with~\cite{Campostrini_90ab} where the $1/N$ corrections to $E_V$ are
computed. Our result is in complete agreement.
\item[3)]  For $r=0$, $d=2$ we get numerically
\be
(E_V)_{3l} \, =\, -0.00599 (N-1)^2 - 0.00727 (N-1)
\ee
which must be compared with the numerical result by
L\"uscher~\cite{Luscher_unpub}:
\be
(E_V)_{3l} \, =\, -0.006 (N-1)^2 - 0.0075 (N-1)
\ee
The two expressions are in reasonable agreement although the coefficient
of $(N-1)$ differs by 3\%, probably due to numerical uncertainties.
\end{itemize}


\section{Long-Distance Quantities} \label{sec4}

\subsection{Correlation Length}   \label{sec4.1}

In this subsection we compute the expected behaviour of the correlation length
in the limit $\beta\to\infty$ in two dimensions using the fact that the model
is
asymptotically free.

In order to do this, one must firstly compute the $\beta$-function for the
lattice
theory. In principle this can be done through a direct lattice computation.
However it is much simpler to take advantage of the fact that this calculation
has
already been done for the continuum theory: indeed the knowledge of the
$\beta$-function in a specific regularization at $n$-loops allows the
determination
of the $\beta$-function at the same number of loops in any other regularization
by simply performing an $(n-1)$-loops computation. Here we will compute
the lattice $\beta$-function at three loops by performing a two-loop
computation on the lattice and using the continuum results by Br\'ezin and
Hikami~\cite{Brezin-Hikami}.

The continuum theory is defined by the bare Hamiltonian
\be
\beta_B H_B = {\beta_B\over 2 } \int {\rm d}x\,
\partial_\mu\bsigma_B\cdot \partial_\mu\bsigma_B
- \beta_B h_B \int {\rm d}x\, \sigma^1_B  \label{continuum-action-bare}
\ee
The theory is ultraviolet divergent and thus it is necessary to
introduce a regularization. Br\'ezin and Hikami~\cite{Brezin-Hikami} use the
dimensional  regularization and renormalize the theory according to the
$\overline{MS}$-scheme at the scale $\mu$.
The renormalized fields and couplings are defined through
\begin{eqnarray}
\beta_B \, &=& \mu^{-\epsilon} \zeta_1^{-1} \beta_r \nonumber \\
\bpi_B  \, &=& {\zeta_2}^{1/2}  \bpi \label{field-redefinitions}\\
h_B \, &=& {\zeta_1\over {\zeta_2}^{1/2}} h \nonumber
\end{eqnarray}
The constants $\zeta_1$ and $\zeta_2$ depend on the the renormalized coupling
$\beta_r$ and on $\epsilon = 2-d$ and are divergent in the limit
$\epsilon\to0$.
They are explicitly given through three loops by
\begin{eqnarray}
\zeta_1(\beta_r)\, &=& 1 - {N-2\over2\pi\epsilon} t \, +\,
\left[{(N-2)\over\epsilon^2} - {N-2\over 2\epsilon}\right] {t^2\over 4\pi^2} \\
&+& \left[ -{(N-2)^3\over\epsilon^3} +{7\over6} {(N-2)^2\over\epsilon^2} -
{N^2-4\over12\epsilon}\right] {t^3\over 8\pi^3} +\, O(t^4) \nonumber \\
\zeta_2(\beta_r) \, &=& 1 - {N-1\over2\pi\epsilon} t \, +\,
{(N-1)(2N-3)\over 2\epsilon^2} {t^2\over 4\pi^2} \\
&+& (N-1)\left[ -{1\over 6 \epsilon^3}(6 N^2 - 19N + 15) +
{(N-2)\over3 \epsilon^2} -
{(N-2)\over 4\epsilon}\right] {t^3\over 8\pi^3} +\, O(t^4) \nonumber
\end{eqnarray}
where
\be
t\, =\, {(4\pi)^{\epsilon/2}\over \Gamma(1-\epsilon/2)} {1\over\beta_r}
\ee
The renormalized Green functions depend now on $\beta$ and $\mu$. However
this two-parameter freedom is only apparent: indeed it is well-known
that each renormalized theory is labelled by a unique dimensional
parameter, the scale $\Lambda_{\overline {MS}}$. Correspondingly all choices
of $\beta_r$ and $\mu$ which satisfy the renormalization-group relation
\be
\mu\, =\, \Lambda_\msbar \left({w_0\over \beta_r}\right)^{w_1/w_0^2}
\exp\left( {\beta_r\over w_0}\right)\,
\exp\left[\int_0^{1/\beta_r} dt\left({1\over W^{\msbar}(1/t)} +
{1\over w_0t^2} -
{w_1\over w_0^2 t}\right)\right] \label{RG-eq}
\ee
define the same theory.
Here $w_0$ and $w_1$ are the first two coefficients of the perturbative
$\beta$-function $W^{\msbar}(\beta)$
\be
W^{\msbar}(\beta)\, =\,
- {w_0\over\beta^2} - {w_1\over\beta^3} -
{w_2^{\msbar} \over\beta^4} -
{w_3^{\msbar} \over\beta^4}
+\, O(\beta^{-5})
\ee
We have not added the superscript $\msbar$ to $w_0$ and $w_1$
because they are universal
in the sense that they do not depend on the renormalization procedure.
They are explicitly given by
\begin{eqnarray}
w_0\, &=&\, {N-2\over 2\pi} \\
w_1\, &=&\, {N-2\over (2\pi)^2}
\end{eqnarray}
On the other hand the other coefficients, $w_2^\msbar$,
$w_3^\msbar$, $\ldots$,  and thus the total $\beta$-function $W^\msbar
(\beta)$,
depend on the regularization. For the
continuum theory in the $\overline{MS}$-scheme we
have~\cite{Brezin-Hikami,Hikami,Wegner_1,Wegner_2}
\begin{eqnarray}
w_2^{\overline{MS}}& =& {1\over4} {N^2-4\over (2\pi)^3} \\
w_3^{\overline{MS}}& =& {N-2\over (2\pi)^4}\left[-{1\over12}(N^2-22N+34)
+{3\over2} \zeta(3) (N-3)\right]
\end{eqnarray}
where $\zeta(3)\approx1.2020569$.

Let us now turn to the lattice theory considering the lattice
$n$-point 1-particle-irreducible (1PI) correlation function.
It can be perturbatively
expanded: the $n$-loop term has the general structure
\be
a^{dim(G)}\, \left[\sum_{k=0}^n\, g_k(ap) \log^k (a^2 h) + X(ap,a^2h)\right]
\ee
where $dim(G)$ is the canonical dimension of the Green function,
$a$ is the lattice spacing and
$X$ vanishes when $a\to0$. As $a\sim \exp(-2\pi\beta/(N-2))$, in the large
$\beta$ limit, $X$ gives non-perturbative (exponentially suppressed)
corrections. Let us now indicate with $\Gamma^{(n)}_{latt}(p_1,\ldots,
p_n;\beta,h;1/a)$ the perturbative 1PI-correlation with $X$ set to zero. As
the lattice is a regularization of the continuum theory
$\Gamma^{(n)}_{latt}(p_1,\ldots, p_n;\beta,h;1/a)$ is the correlation
in a continuum theory renormalized
at the scale $1/a$ and thus, modulo a finite renormalization, it must
be equal to the same Green function in any other regularization.
If we indicate with $\Gamma^{(n)}_{\overline{MS}} (p_1,\ldots,p_n;\beta,h;\mu)$
the $\overline{MS}$-renormalized correlation,
the general results of~\cite{Brezin-LeGuillou-ZinnJustin} imply that there
exist finite constants
$Z_1(\beta,\mu a)$ and $Z_2(\beta, \mu a)$ such that
\be
\Gamma^{(n)}_{latt} (p_1,\ldots,p_n;\beta,h;1/a)\, =\, Z_2^{n/2}\,
\Gamma^{(n)}_{\overline{MS}} (p_1,\ldots,p_n;Z_1^{-1}\beta,Z_1 Z_2^{-1/2} h;\mu
)
\label{latticeMS}
\ee
Both functions $\Gamma^{(n)}_{latt} (p_1,\ldots,p_n;\beta,h;1/a)$ and
$\Gamma^{(n)}_{\overline{MS}} (p_1,\ldots,p_n;\beta,h;\mu )$ satisfy a
renorma\-li\-za\-tion-group equation. For the lattice Green function it has the
form~\cite{Brezin_76}
\be
\left[-a{\partial\over\partial a}+\, W^{latt}(\beta)
{\partial\over\partial\beta^{-1}}-\, {n\over2}\,
\gamma^{latt}(\beta)+\, \left({1\over2}\gamma^{latt}(\beta)+
\,\beta W^{latt}(\beta)\right)\,
h{\partial\over\partial h}\right]\, \Gamma^{(n)}_{latt}\, =\, 0
\label{RGGreen}
\ee
and for $\Gamma^{(n)}_\msbar$ the same holds using $W^\msbar(\beta)$ and
$\gamma^\msbar(\beta)$. Here $W(\beta)$ stands for the $\beta$-function
we introduced above, while $\gamma(\beta)$ is the anomalous dimension of the
field $\bpi$. Also $\gamma(\beta)$ depends on the regularization and
for this reason we have added different superscripts.

Let us compute the relation between the $\beta$-functions and the
anomalous dimension in the continuum in the $\msbar$-scheme and on the lattice.
Substituting (\ref{latticeMS}) into (\ref{RGGreen}) one immediately identifies:
\begin{eqnarray}
W^\msbar(Z_1^{-1} \beta) &=& W^{latt} (\beta)\left( Z_1 + {1\over\beta}
{\partial Z_1\over\partial\beta^{-1}}\right) \label{eq2.74}\\
\gamma^\msbar(Z_1^{-1} \beta) &=& \gamma^{latt}(\beta) - \,
W^{latt} (\beta) {1\over Z_2} \,{\partial Z_2\over\partial\beta^{-1}}
\label{eq2.75}
\end{eqnarray}

These relations allow us to compute $W^{latt}(\beta)$ and
$\gamma^{latt}(\beta)$ at $n$-loops in terms of the corresponding
quantities in the continuum at $n$-loops and of the renormalization
constants $Z_1$ and $Z_2$ at $(n-1)$ loops.

Thus our main task is the computation of $Z_1$ and $Z_2$.
It is easily seen that, in order to compute them, it
is enough to match the 1PI two-point function
$\Gamma^{(2)}_{ij} (p) = \delta_{ij}\beta(p^2 + h) - \Pi_{ij}(p)$.
At one loop on the lattice
the vacuum polarization $\Pi_{ij}(p) = \Pi(p)\,
\delta_{ij}$ is given by (see Figure~\ref{fig2})
\begin{figure}
\vspace*{0cm} \hspace*{-0cm}
\begin{center}
\epsfxsize = 0.8\textwidth
\leavevmode\epsffile{3loops2.ps}  
\end{center}
\vspace*{0cm}
\caption{
  Feynman graphs relevant to the evaluation of the two-point function
 respectively (a) at the tree-level (b) at one-loop (c) at two-loop.
  The cross represents the vertices coming
  from the measure term in $H^{eff}$.}
\label{fig2}
\end{figure}
\begin{eqnarray}
\Pi^{latt}(p) &=& {\hatp^2\over 4} + r\,(N+1) {\hatp^2 \over 4}
(1 - hI(h))
- \left[
\hatp^2 + {h\over 2} (N-1) + {\hatp^2\over 4} h \right] I(h)  \nonumber  \\
&=& {p^2\over 4} + r\,(N+1) {p^2 \over 4} + {1\over 4\pi}
\left[
p^2 + {h\over 2} (N-1)\right] \log{ha^2 \over 32} \, + \, O(a) \;.
\label{propagatore}
\end{eqnarray}
For the continuum theory  in the
$\normalmsbar$-scheme at the scale $\mu$ we get instead
\be
\Pi^\msbar (p) = {1\over 4\pi} \left[
p^2 + {h\over 2} (N-1)\right] \log{h\over \mu^2} \;
\ee
By comparing the two expressions we obtain the one-loop results
\begin{eqnarray}
Z_1 &=& 1 - {N-2\over 4 \pi \beta} \log {\mu^2 a^2 \over 32} + {1\over 4 \beta}
\left (1 + r (N+1)\right) +\; O(1/\beta^2) \nonumber \\
Z_2 &=& 1 -{N-1\over 4\pi\beta}  \log  {\mu^2 a^2 \over 32}+\; O(1/\beta^2)\; .
\label{Z_cont_latt}
\end{eqnarray}
Let us now compute the renormalization-group relation between $a$ and $\beta$.
Starting from (\ref{RG-eq}) and using (\ref{Z_cont_latt}) we get
\begin{eqnarray}
{1\over a}\, &=& \, {\Lambda_{\overline{MS}}\over \mu a}
\left({(N-2)Z_1\over 2\pi\beta}\right)^{1/(N-2)}
\exp\left( {2\pi\beta\over (N-2)Z_1}\right)\, \left(1+ O(\beta^{-1})
\right) \\
&=& \, \Lambda_{latt}
\left({N-2\over 2\pi\beta}\right)^{1/(N-2)}
\exp\left( {2\pi\beta\over N-2}\right)\, \left(1+ O(\beta^{-1})
\right) \label{one-over-a}
\end{eqnarray}
where
\be
\Lambda_{latt} = \Lambda_\msbar\, 2^{-5/ 2} \exp
\left[-{\pi\over 2(N-2)} \left(1 +  r(N+1)\right) \right]
\ee
As special cases we have
\begin{eqnarray}
\Lambda_{latt,N-vector} &=& \Lambda_\msbar \,  2^{- 5/ 2}
\exp \left[- {\pi\over 2(N-2)}\right] \\
\Lambda_{latt,RP^{N-1}} &=& \Lambda_\msbar \,  2^{- 5/ 2}
\exp \left[-  {\pi\over 2(N-2)}\right] \exp \left[- { \pi\over 2}\,{ N+1\over
N-2}\right] \end{eqnarray}
(where the first formula had been obtained in~\cite{Parisi_80})
and thus
\be
\Lambda_{latt,RP^{N-1}} = \Lambda_{latt,N-vector}\, \exp \left[- { \pi\over
2}\,
   {N+1\over N-2} \right] \label{RP-to-ON-Lambda}
\ee
which was obtained for the case $N=3$ in~\cite{Sinclair_82}.
We can also compare with~\cite{Caselle-Gliozzi-Megna}, by denoting with the
superscript $CGM$ their definitions.
As $\beta = \beta_F^{CGM} + 8 \beta_A^{CGM}/3$
and $r\ =\ 8 \beta_A^{CGM}/ 3\beta$
we find that their Eq. (15) agrees with ours for $N=4$. Moreover our
result  (\ref{RP-to-ON-Lambda}) agrees in the limit $N\to\infty$ with
the result of \cite{Magnoli-Ravanini}.

We want now to derive
the first analytic correction in (\ref{one-over-a}) which requires the
computation
at the two-loop level of the constants $Z_1$ and $Z_2$. At two-loops in the
continuum
the self-energy is given by
\begin{eqnarray}
 {1\over4\pi^2}  & &\! \left[ -{N-1\over8} \log^2 {h \over \mu^2} -
{5N-11\over8}
\log {h \over \mu^2} +{N-2\over4} - (2\pi)^2 (N-2) R\right]\, p^2 \nonumber \\
- {N-1\over 4\pi^2}& &\! \left[{3N-5\over32} \log^2 {h \over \mu^2}
+{N-3\over16}
\log {h \over \mu^2}\right]\, h
\end{eqnarray}
while on the lattice we get
\begin{eqnarray}
&& {1\over4\pi^2}\left[ -{N-1\over8} \log^2 {h a^2\over 32} - {5N-11\over8}
\log {h a^2\over 32} +{11-3N\over24}\pi^2 \right. \nonumber \\
&& \left. + {\pi\over4} + (2\pi)^2 (N-2) (G_1-R)
\right]\, p^2 \nonumber\\
&-& {N-1\over4\pi^2}\left[{3N-5\over32} \log^2 {h a^2\over 32}
+\left ( {N-3\over16} -{\pi\over8} \right)
\log {h a^2\over 32} - {\pi\over8}\right]\, h \nonumber \\
&+&{(N+1)\over 8} r \left\{
p^2\left[ {1\over 12} +{1\over 2\pi}\right]
+ {N-1\over 4\pi} h \left[ \log {h a^2\over 32} + 1 \right] \right\} \nonumber
\\
& & \quad +{(N+1) (N+2)\over 16}\, r^2 p^2 - {(N+1)(N+3)\over 32} \, s\, p^2
\end{eqnarray}
If we now indicate with $Z_{12}$ and $Z_{22}$ the terms in $Z_1$ and $Z_2$
proportional to $1/\beta^2$ we get
\begin{eqnarray}
Z_{12} &=& {(N-2)^2\over 16\pi^2} \log^2 {\mu^2 a^2\over 32}
- {N-2\over 8\pi} \left( 1 + {1\over\pi} \right) \log {\mu^2 a^2\over 32}
\nonumber  \\
&+& {(N-2)\over 4\pi^2} \left[ 4\pi^2 G_1 - {1\over4} - {\pi^2\over8}\right]
+{11\over96} \nonumber\\
&-& {r\over 8\pi} (N+1)(N-2) \log {\mu^2 a^2\over 32} + {13\over96}(N+1)r \\
&+& {r^2\over16}(N+1)(2N+3) - {s\over 32}(N+1)(N+3)
\nonumber \\
Z_{22} &=& {1\over32\pi^2} (N-1)(2N-3) \log^2 {\mu^2 a^2\over 32}
- {N-1\over16\pi}  [1+r(N+1)] \log {\mu^2 a^2\over 32}  \nonumber
\end{eqnarray}
Using formula (\ref{eq2.74}) we can now compute
the three-loop coefficient $w_2^{latt}$
for the lattice $\beta$-function. We get
\begin{eqnarray}
w^{latt}_2 &=& {N-2\over (2\pi)^3}\left[\left({1\over2} + {\pi^2\over8} -
4\pi^2 G_1
\right)
(N-2) + 1 + {\pi\over2} - {5\pi^2\over24}\right]  \\
& & \quad + {1\over2\pi} (N+1)(N-2)\left[\left({1\over8\pi} -
{1\over96}\right)\ r
-{N+2\over16}\ r^2 + {N+3\over 32} s\right] \nonumber
\end{eqnarray}
For $r=s=0$ our result agrees with that obtained by Falcioni and
Treves~\cite{Falcioni-Treves},  as their integral $G_2$ can be computed exactly
(see
Appendix A) and is given by
\be
 \label{eq2.87}
G_2\ =\,\lim_{h\to 0} \int {{\rm d}^2p \over (2\pi)^2} {{\rm d}^2 q \over
(2\pi)^2}
{\sum_\mu (2 \sin p_\mu  \sin q_\mu + \sin^2 p_\mu ) \over
(\widehat{(p+q)^2}+h) (\hat{p}^2+h) (\hat{q}^2+h)} \, =\, {1\over 48}   \;.
\ee
Then let us consider the renormalization-group relation
\begin{eqnarray}
{1\over a}& =& \Lambda_{latt} \left({w_0\over \beta}\right)^{w_1/w_0^2}
\exp\left( {\beta\over w_0}\right)\,
\exp\left[\int_0^{1/\beta} dt\left({1\over W^{latt}(1/t)} +
{1\over w_0t^2} -
{w_1\over w_0^2 t}\right)\right] \\
&=& \Lambda_{latt} \left({N-2\over 2\pi\beta}\right)^{1/(N-2)}
\exp\left( {2\pi\beta\over N-2}\right)
\left[ 1 + \sum_{n=1}^\infty {a_n\over\beta^n} \right]
\label{RG-eq-latt}
\end{eqnarray}
Using $w_2^{latt}$ we can now compute the first analytic correction to the
previous formula. The coefficient $a_1$ is given by
\begin{eqnarray}
a_1 &=& {1\over N-2} \left[\left({1\over4\pi} +
{\pi\over16} - 2\pi G_1\right) (N-2)+ {1\over4} -
{5\pi\over48}\right]
\nonumber \\
&+&  2\pi\, {N+1\over N-2}\left[ \left({1\over 8\pi} - {1\over 96}
\right) r -
{N+2\over 16} r^2 + {N+3\over32} s\right]
\end{eqnarray}
Using this expression and the analytic result
\cite{Hasenfratz-Niedermayer_1,Hasenfratz-Niedermayer_2,Hasenfratz-Niedermayer_3}
\be
{m\over\Lambda_\msbar}\, =\, \left( {8\over e} \right)^{1/(N-2)} \,
{1\over \Gamma(1+1/(N-2))}  \;,
\ee
one can compute the exponential correlation length. We get
\begin{eqnarray}
\xi^{(exp)}_V (\beta) &=& (ma)^{-1} \nonumber \\[3mm]
&=& C_{\xi_V}
\left({N-2\over 2\pi\beta}\right)^{1/(N-2)}
\exp\left( {2\pi\beta\over N-2}\right) \left(1 +
\sum_{n=1}^\infty{a_n\over\beta^n} \right)
\label{eq2.90}
\end{eqnarray}
where
\begin{eqnarray}
C_{\xi_V} &=& \left( {e\over 8} \right)^{1/(N-2)}
\Gamma\left(1+{1\over N-2}\right) 2^{-5/2}
\nonumber \\
&& \exp\left[-{\pi\over 2(N-2)}\left(1+r(N+1)\right)\right]
\end{eqnarray}
This expression can be checked in the limit $N\to\infty$ with
$\widetilde{\beta} \equiv \beta/N$ fixed
and $s=0$, by comparing to the results in Appendix \ref{secB}
for the large-$N$ limit of the
mixed isovector-isotensor action \reff{eq1.1}.
Indeed, \reff{eq2.90} agrees with \reff{eqC.19}.

Let us notice that this expression for $\xi_V^{(exp)}$
is not valid for $RP^{N-1}$ models, where of course $\xi_V^{(exp)}=0$.
In that case one must consider $\xi_T^{(exp)}$,
i.e. the correlation length in the tensor channel. The dependence on $\beta$
is again given by the previous formula, but the non-perturbative constant
will be different. If we accept the standard scenario which postulates that
the $RP^{N-1}$ models are simply $O(N)$ models restricted to the
$Z_2$-invariant (isotensor) sector,
and take into account that the mass in the tensor channel in an $O(N)$ model
is exactly twice
the mass in the vector channel since no bound states exist, we then get
an expression for $\xi^{(exp)}_T$
which is exactly one-half of the previous one.\footnote{
   However, in~\cite{CEPS_93} we shall present heuristic and numerical evidence
   that this standard scenario is {\em not}\/ true,
   and that the equality $\xi_V^{(exp)}/\xi_T^{(exp)} = 2$
   is valid only to the right of a certain curve in the
   $(\beta_V,\beta_T)$-plane which tends to
   $\beta_V = \beta_{c,Ising} = {1 \over 2} \log(1 + \sqrt{2}) \approx
0.440687$
   as $\beta_T \to \infty$.
}
As the difference between the two cases is a simple constant factor,
in the following we will refer to $\xi^{(exp)}_V$ also for $RP^{N-1}$ models
in order to treat $O(N)$ models and $RP^{N-1}$ models together.

For the case $N=3$, we get the explicit formula
\begin{eqnarray}
\xi^{(exp)}_V (\beta) &=& {e\over 32\sqrt{2}}
{e^{2\pi\beta}\over 2\pi\beta} \exp \left[ -{\pi\over2}
\left(1 + 4r\right)\right]
  \\
&&\left\{1 -{1\over\beta}\left[2\pi G_1 - {1\over4\pi} - {1\over4} +
{\pi\over24} - \left(1 - {\pi\over12}\right)r + {5\pi\over2}r^2
   - {3 \pi\over 2} s \right]\right\}
\nonumber
\end{eqnarray}
Numerically the analytic non-universal correction is given by
\be
1 - {1\over\beta}\left[0.091 - 0.738\, r + 7.854\, r^2 - 4.712 \, s \right]
\ee
Let us notice that for $r=1$, $s=0$, i.e. in the $RP^2$ model, the correction
is extremely large and indeed the $1/\beta$ term is larger than 1 at the values
of
$\beta$ ($\ltapprox 5.6$) which can be investigated by present-day algorithms.
Roughly speaking, the perturbative result predicts its own unreliability
for $\beta \ltapprox 15$, that is up to enormous correlation lengths.
This provides a plausible explanation of the observed discrepancies
\cite{Sinclair_82,CEPS_LAT92} between the two-loop perturbative predictions
for the $RP^2$ model and the Monte Carlo simulations.

Completely analogous results can be derived for $N=4$.
In this case the analytic correction is given by
\begin{eqnarray}
&&1 -{1\over\beta}\left[2\pi G_1 - {1\over 4\pi} - {1\over8} - {\pi\over96}  -
{5\over8}\left(1 - {\pi\over12}\right)
r + {15\pi\over8}r^2 - {35\pi\over 32} s  \right]
\nonumber \\
&=&1 - {1\over\beta}\left[0.053 - 0.461\, r + 5.890\, r^2 - 3.436 \, s \right]
\end{eqnarray}
If $r=1$ and $s=0$ we get again a large correction for $\beta\approx 6$
which is the value used in the Monte Carlo simulation of $RP^3$.

The situation improves when $N$ is increased.
As $N \to \infty$ at fixed $\beta/N$,
the analytic correction term becomes
\be
  1 - {N \over {\beta}}
                 \left[ {\pi\over 8} r^2 - {\pi\over 16} s \right]    \;,
\label{correzione-N-infinito}
\ee
Now, for the case $r=1, s=0$, using the results of Appendix \ref{secB}, we find
that $\xi=10$ (respectively 100) corresponds to $\beta/N=1.047$ (resp. 1.341).
Thus in the large-$N$ limit, in the region accessible to Monte Carlo
computations, the correction is of order 30-40\% which is still quite large.
Notice that if we were using the two-loop approximation we would get
for $\beta/N=1.047$ (respectively 1.341) $\xi=26$ (resp. 168), while with the
three-loop approximation we would have $\xi=17$ (resp. 119). Thus in
simulations of $RP^{N-1}$ for large $N$ we expect a discrepancy by
a factor of two if we use the two-loop formula while the three-loop
approximation reduces the error to 20-40\%.

However, the fact that a large analytic correction is present at small $N$
could nevertheless have been
suspected from the $1/N$ expansion together with some knowledge
of the form of the $N$-dependence in the coefficients.
Indeed, from (\ref{correzione-N-infinito}), setting $s=0$,
keeping into account the
structure of the vertex proportional to $r$, we obtain for finite $N$
\be
1 + \,{1\over (N-2)\beta}\,\left( - {\pi\over8} r^2 (N+1)(N+\alpha_1)\,
+ \alpha_2 (N+1) r + \alpha_3 N + \alpha_4\right) + \, O(1/\beta^2)
\ee
where $\alpha_1,\ldots,\alpha_4$ are constants to be determined.
For $N=3$ it gives
\begin{eqnarray}
1 &+& \, {1\over\beta}\, \left( - {\pi\over2} r^2 (3 + \alpha_1) +
\, 4 \alpha_2 r\, + 3 \alpha_3 + \alpha_4\right)\approx\, \nonumber \\
1 &+& \, {1\over\beta}\, \left( (- 4.712 + 1.571 \alpha_1) r^2
+ \, 4 \alpha_2 r\, + 3 \alpha_3 + \alpha_4\right)
\end{eqnarray}
Thus, unless some cancellation happens among different terms, large corrections
must be expected. For instance, if we set all the unknown constants to
zero we get a correction which has the correct order of magnitude
(within a factor of 2).
This simple reasoning tells us also
that at the next order the largest corrections (when $r \sim 1$)
for the theory defined by (\ref{eq1.1})
are presumably those coming
from the $r^3$ contribution to the four-loop $\beta$-function which is the
only one surviving in the large-N limit.
We have thus undertaken
the task of computing this term, as well as that proportional to $r^2$,
in order
to check if our assumption is correct.
We must thus compute the lattice self-energy
at three loops. The relevant graphs are reported in Figure~\ref{fig3}.
\begin{figure}
\vspace*{0cm} \hspace*{-0cm}
\begin{center}
\epsfxsize = 0.8\textwidth
\leavevmode\epsffile{3loops3.ps}  
\end{center}
\vspace*{0cm}
\caption{
Graphs contributing terms proportional to $r^2$ and $r^3$ to the self-energy
to three loops.}
\label{fig3}
\end{figure}
We get
\begin{eqnarray}
&&r^3 p^2 (N+1)\left[ {(N+1)^2\over32}\, +\, {3\over64}(N+1)\, +\, {3\over16}
J (N+3) \right] \nonumber \\
&+& r^2 p^2 (N+1)\left[-{N+9\over 128} +\, {N+28\over 48} J+\,
{3N+5\over128\pi}
-{L_1\over2} - {N-2\over8}(L_2+4L_3) \right]  \nonumber \\
&+& r^2 h (N^2-1)\left[{2N+3\over128\pi}\log{ha^2\over32}+\, {5N+7\over256\pi}
-\, {L_1\over4}\right]
\end{eqnarray}
Thus, if we indicate with $Z_{13}$ and with $Z_{23}$ the contributions to
$Z_1$ and $Z_2$ proportional to $1/\beta^3$ we have
\begin{eqnarray}
Z_{13} &=& {5 r^3\over 64} (N+2)(N+1)^2\, +\, {3 r^3\over16} (N+1)(N+3)J
    \nonumber \\
   &+& r^2 (N+1) \left[- {1\over64\pi} (5N+7)(N-2) \log{\mu^2 a^2\over32}
  +\, {1\over 384} (29N+17) \right. \nonumber  \\
   && \qquad +\, \left. {J\over48}(N+28)\, +\, {N-2\over8} (4 L_1 - L_2 - 4
L_3)
      \right] \\
Z_{23} &=& r^2 (N^2-1)\, \left[-{1\over64\pi} (2N+3) \log{\mu^2 a^2\over32} +\,
{L_1\over2}\right]
\end{eqnarray}
Correspondingly we get for the terms proportional to $r^2$ and $r^3$ in the
four-loop $\beta$-function
\begin{eqnarray}
w^{latt}_3 &=& {(N-2)(N+1)\over 2\pi}\left[
- {1\over 32} (N+1)(2N+5) r^3 - {3\over 8} (N+3) J r^3 \right.\nonumber \\
&&\qquad + {N+9\over 64}r^2\, -\, {1\over4}(N-2)(4L_1 - L_2 - 4 L_3) r^2
\nonumber \\
&& \qquad \left. - {J\over 24}(N+28)r^2 + {1\over 32\pi} (N+1) r^2\right]
\end{eqnarray}
Thus
for the terms in $a_2$ (defined in \reff{eq2.90})
proportional to $r^2$, $r^3$ and $r^4$ we have
\begin{eqnarray}
a_2 &=& {\pi^2 r^4\over 128} {(N+1)^2 (N+2)^2\over (N-2)^2} \nonumber \\
&-&
{\pi\over 16} r^3{N+1\over (N-2)^2}\left[ (N+1)
\left(N^2 + N - 4 - {\pi\over 24} (N+2)
\right) + 6 (N+3) (N-2) J\right] \nonumber \\
&+& r^2 {N+1\over(N-2)^2}\, \left[-{\pi\over4}(N-2)^2 (4 L_1 - L_2 - 4 L_3)
-{\pi\over24} J (N+28)(N-2) \right. \nonumber \\
&& \qquad +{\pi^2\over4}(N^2-4)G_1 \,
+\, {\pi^2\over 4608}(-36 N^2 + 61 N + 265) \nonumber \\
&& \qquad \left. + {1\over32}(2N^2-5) \, +\, {\pi\over 192}(3 N^2 + 14 N -67)
\right]
\end{eqnarray}
Numerically for $N=3$ we get
\be
a_2 =\, 2.66 \, r^2 -26.94 \, r^3  + 30.84 \, r^4
\ee
while for $N=4$ we have
\be
a_2 =\, 1.64 \, r^2 -21.49 \, r^3  + 17.35 \, r^4
\ee
{}From these formulae one sees that, if $r\sim 1$,
the contributions proportional to
$r^3$ and $r^4$ are much larger than those proportional to $r^2$. However
notice that the $r^3$ and $r^4$ terms partially cancel themselves
leaving a correction which is only slightly larger than the $r^2$
contribution. In any case we expect the $r$-term we have not computed
to be at most of the same magnitude of the $r^2$-term, while the
good agreement between simulations and the theoretical prediction
of $\xi$ for the $O(N)$-model with standard action tells us that
the contribution independent from $r$ is probably much smaller.
Thus for $RP^2$, $r=1$, the analytic corrections are estimated as
\be
1 -\,  {7.21\over \beta} \, +\, {6.56\pm 2.7\over \beta^2} \, +\, O(1/\beta^3)
\ee
which, even though the new term has positive sign,
are still larger than 1 at the values of $\beta$ used in
Monte Carlo simulations.
For $RP^3$ we get instead
\be
1 -\, {5.48\over \beta} \, -\, {2.50\pm 1.6\over \beta^2} \, +\, O(1/\beta^3)
\ee
which vanishes for $\beta\approx 6$ which is the region where
Monte Carlo simulations are made.

For large-$N$ our result reduces to
\be
a_2 = \left({\pi^2\over128}r^4 \, -\, {\pi\over16}r^3\right)\, N^2
\ee
which agrees with the direct expansion of \reff{eqC.19}.

\subsection{Long-distance Quantities: Susceptibilities}   \label{sec2.5}

In this subsection we compute the expected behaviour of the
susceptibilities.

Let us begin with the vector susceptibility
\be
\chi_V \, =\, \sum_x \<\bsigma_0\bsigma_x\>
\ee
We need to compute the anomalous dimension $\gamma^{latt}(\beta)$
of the field $\bpi$. In the continuum in the $\msbar$-renormalization
$\gamma^\msbar(\beta)$ has the form
\be
\gamma^\msbar(\beta)\, =\, {\gamma_0\over\beta}+\,
{\gamma_1^\msbar\over\beta^2}+\,
{\gamma_2^\msbar\over\beta^3}+\,
{\gamma_3^\msbar\over\beta^4}+\, O(1/\beta^5)
\ee
with~\cite{Brezin-Hikami,Hikami,Wegner_1,Wegner_2}
\begin{eqnarray}
\gamma_0 &=& {N-1\over 2\pi} \\
\gamma_1^\msbar &=& 0 \\
\gamma_2^\msbar &=& {3\over32\pi^3}(N-1)(N-2) \\
\gamma_3^\msbar &=& {1\over 192\pi^4}(N-1)(N-2)[4(5-N)+3(3-N)\zeta(3)]
\end{eqnarray}
where $\zeta(3)\approx1.2020569$.

As the first coefficient $\gamma_0$ is universal,
i.e. regularization independent, we have dropped the superscript $\msbar$.
Using (\ref{eq2.75}) we get
\begin{eqnarray}
\gamma_1^{latt} &=& \, {N-1\over8\pi} (1 + r(N+1))\\
\gamma_2^{latt} &=& \, {N-1\over2\pi}\left[ {N-2\over4\pi^2}
\left(4 \pi^2 G_1 + {1\over2} - {\pi^2\over8} \right)\, +\, {11\over96}
\right. \nonumber \\
&& \quad \left.+{13\over96}(N+1)r\, +{r^2\over16}(N+1)(2N+3) -
{s\over32} (N+1)(N+3) \right]
\end{eqnarray}
while for the $r^2$ and $r^3$ contributions to $\gamma^{latt}_3$ we have
\begin{eqnarray}
\gamma^{latt}_3 &=& {N^2-1\over2\pi}r^3\left[{5\over64}(N+2)(N+1)
+{3\over16}(N+3)J\right]  \\
&+& {N^2-1\over16\pi}r^2\left[ {1\over48}(29N+17)+
{J\over6}(N+28) - (N-2)(8 L_1 + L_2 + 4 L_3)\right] \nonumber
\end{eqnarray}
We can now derive the general expression for the vector susceptibility.
Starting from the renormalization-group equation~\cite{Brezin_76}
\be
\left[-a{\partial\over\partial a}+\,
W^{latt}(\beta){\partial\over\partial\beta^{-1}}+\,
\gamma^{latt}(\beta) \right]\, \chi_V \, =\, 0
\ee
we get
\begin{eqnarray}
\chi_V &=& C_\chi e^{2\beta/w_0}
\left( {w_0\over \beta}\right)^{2 w_1/w_0^2+\gamma_0/w_0} \nonumber \\
&& \exp\left[\int_0^{1/\beta} dx\, \left({2\over W^{latt}(1/x)} +
\, {2\over w_0 x^2} -\, {2w_1\over w_0^2 x}-\,
{\gamma^{latt}(1/x)\over W^{latt}(1/x)}
- \, {\gamma_0\over w_0 x}\right)\right]
\nonumber \\
&=& C_\chi e^{4\pi\beta/(N-2)}
\left({2\pi\beta\over (N-2)}\right)^{-(N+1)/(N-2)}\,
\left\{1 + {b_1\over\beta}+\, {b_2\over \beta^2} + O(1/\beta^3)\right\}
\label{eq2.131}
\end{eqnarray}
The (non-universal) constant $C_\chi$ cannot be computed in perturbation
theory.
However the ratio of $C_\chi$ in different models in the same
universality class is simply given by the square of the ratio of
the $\Lambda$-parameters.
Thus
\be
{C_\chi(r)\over C_\chi(r=0)}\, =\, \exp\left(-\pi r{N+1\over N-2}\right)
\ee
We can estimate $C_\chi$ in the large $N$-limit.
Using the previous relation and the $1/N$ results of
Ref. \cite{Campostrini_90ab}
we obtain the following estimate:
\begin{eqnarray}
C_\chi &=& {\pi\over 16}\, \left[1 + {1\over N} (4 + 3 \gamma_C
-\pi - 3 \gamma_E - 7 \log 2) +\, O(1/N^2)\right]
\exp\left(-\pi r{N+1\over N-2}\right) \nonumber \\
& \approx & 0.196 \, ( 1 - 4.267/N + \, O(1/N^2))\,
\exp\left(-\pi r{N+1\over N-2}
\right)
\end{eqnarray}
where $\gamma_E\approx 0.5772157$ is Euler's constant and
\be
\gamma_C \, =\, \log\left({\Gamma(1/3)\Gamma(7/6)\over\Gamma(2/3)\Gamma(5/6)}
\right)\, \approx \, 0.4861007
\ee
In order to compute the analytic corrections let us rewrite
\be
{\chi_V }\, =\, R_V \xi_V^2\,
\left({2\pi\beta\over N-2}\right)^{-(N-1)/(N-2)} \left(1 + \,
\sum_{n=1}^\infty  {c_n\over \beta^n} \right)
\ee
where $R_V$ is a non-perturbative universal quantity.

Using the computed values of $\gamma^{latt}(\beta)$ we can estimate
$c_1$ and $c_2$ and give the contributions to $c_3$ proportional to
$r^2$ and $r^3$. We have
\begin{eqnarray}
c_1 &=& {1\over 4\pi}\,{N-1\over N-2} [\pi-2 +\pi r(N+1)] \\
c_2 &=& {N-1\over (N-2)^2}\, \left[ - {1\over 96} (3 N^2 - 23 N + 31)
       +{N-1\over 8\pi^2} - {2N-3\over 8\pi} + (N-2)^2 G_1\right] \nonumber \\
    && + {N^2-1\over (N-2)^2}\left[{13N-20\over96}\,r-\, {2N-3\over8\pi}
       \, r +\, {r^2\over32} (4 N^2 - N -11)\right] \nonumber \\
    && - {(N^2-1)(N+3) \over N-2} {s\over 32}
\end{eqnarray}
while for the terms proportional to $r^2$ and $r^3$ in $c_3$ we have
\begin{eqnarray}
c_3 &=& {N^2-1\over (N-2)^3}\, r^3\left[{1\over 384}
        (30 N^4 - 19 N^3 - 173 N^2 + 83 N + 207)\, +\,
        {3\over 16} J(N-2)^2 (N+3) \right] \nonumber \\
    && + {N^2-1\over (N-2)^3}\, r^2\left[{1\over 384}
        (29 N^3 - 77 N^2 + 35 N +15) \right. \nonumber \\
    && - {1\over 64\pi} (10 N^3 - 19 N^2 - 20 N + 39)  \nonumber \\
    && \left. + {J\over 48} (N-2)^2(N+28)-\, {1\over8}
       (N-2)^3 (L_2 + 4 L_3)\right]
\end{eqnarray}
Using now the expansion for the correlation length we can then compute
$b_1$ and the contributions to $b_2$ proportional to
$r^2$, $r^3$ and $r^4$.
We get for $b_1$
\begin{eqnarray}
b_1\,
&= &{1\over 2\pi(N-2)} \left[-1 +{\pi\over2}(N+1) +
{\pi^2\over4}\left(N-{11\over3}\right) - 8\pi^2 (N-2)G_1 \right.
\nonumber \\
&&  +  (N+1){\pi\over2}r\left(N+1-{\pi\over6}\right) - {\pi^2\over2}r^2
(N+1)(N+2) \nonumber \\
&&  \left. + {\pi^2\over4} s (N+1)(N+3)\right]
\end{eqnarray}
%
%
while for the terms in $b_2$ proportional to $r^2$, $r^3$, $r^4$ we get
\begin{eqnarray}
b_2 &=& {\pi^2 r^4\over 32} {(N+1)^2 (N+2)^2\over (N-2)^2} \nonumber \\
&-&
{\pi\over 16} r^3{N+1\over (N-2)^2}\left[ (N+1)
\left(3 N^2 + 4 N - 8 - {\pi\over 6} (N+2)
\right) + 12 (N+3) (N-2) J\right] \nonumber \\
&+& r^2 {N+1\over(N-2)^2}\, \left[-{\pi\over2}(N-2)^2 (4 L_1 - L_2 - 4 L_3)
-{\pi\over12} J (N+28)(N-2) \right. \nonumber \\
&& \qquad
+\, \pi^2 (N^2-4) G_1 +\, {\pi^2\over 1152}(-36 N^2 + 61 N + 265) \nonumber \\
&& \qquad \left. + {1\over32}(4N^3 + 5 N^2 - 4 N - 1) \,
+\, {\pi\over 96}(-4 N^2 + N - 67)
\right]
\end{eqnarray}
The same discussion we have presented for the correlation length
applies to $\chi_V$. For instance for $N=3$
\begin{eqnarray}
b_1 &=& -0.001\, +\, 3.476\, r -\, 15.708\, r^2 + \, 9.425\, s\\
b_2 &=& 12.907\,r^2\, -\, 96.890\, r^3 + 123.370\, r^4 +\, O(r,1)
\end{eqnarray}
We immediately notice that the coefficient of $r^2$ in $b_1$ is extremely
large. Thus for the model defined in \reff{eq1.1} one expects to be able to
see asymptotic scaling only for small values of $r$. If we require
$|b_1|\leq 1$ which roughly corresponds to a 50\% correction in the region
where $\xi_V\approx$ 50--100 (here $\beta\approx 2.00$, see \cite{CEPS_93})
we find $-0.16\leq r \leq 0.39$  which is a small wedge
around the $\beta_T=0$ axis.
The situation improves for
larger values of $N$. In the limit $N\to\infty$, $\beta/N$
fixed, we get
\begin{eqnarray}
b_1 &=& {N\over8}\left[2r-\, 2\pi \, r^2\, + \pi  s\right] \\
b_2 &=& {N^2\over 32} \left[4 \, r^2\, - 6\pi r^3 +\,
\pi^2 r^4\right]
\end{eqnarray}
which agree  for $s=0$ with the direct expansion of
\reff{eqC.17} in Appendix \ref{secB}.
Notice that no term proportional to $r^1$ and $r^0$ survives in $b_2$
in the large-$N$
limit so that our approximate formula for $b_2$ becomes more accurate
for large values of $N$. For $N=\infty$,
for the action \reff{eq1.1}, we can compute exactly the expected
discrepancy between the exact value of $\chi_V$ and the two-loop,
three-loop and four-loop  approximations. For $0\leq r \leq 1$, at a
correlation length of 10, we find that the discrepancy varies from
1\% to a factor of four if one uses the two-loop formula, from 1\% to a factor
of two with the three-loop approximation while the four-loop result
differs at most of 48\%. At a correlation length of 100 the situation is better
with a discrepancy by a factor of two, of 27\% and 9\% at most in the
three cases.

In order to observe asymptotic scaling it is better to consider
$\chi_V/\xi^2$ instead of $\chi_V$.
Indeed in this case the corrections are significantly smaller and
moreover an additional term is available.
For $N=3$ we get
\begin{eqnarray}
c_1 &=& 0.182 + 2.000 \, r \\
c_2 &=& 0.133 +\, 0.628\, r +\, 5.500\, r^2 - 1.500\, s\\
c_3 &=& 2.034 \, r^2 +\, 18.230 \, r^3 +\, O(r,1)
\end{eqnarray}
while for large $N$ we get
\begin{eqnarray}
c_1 &=& {r\over 4}N \\
c_2 &=& {r^2\over8} N^2 - {s\over 32} N^2\\
c_3 &=& {5 r^3\over 64} N^3
\end{eqnarray}
which agrees (for $s=0$)
with the direct expansion of the formulae of Appendix \ref{secB}.

Let us notice that the previous discussion for $\chi_V$ does not apply
to $RP^{N-1}$ models. Indeed $\chi_V$ is not invariant under
$Z_2$-gauge transformations and it is easy to see that $\chi_V=1$
identically. For this class of models the natural field
is the spin-two operator $\bsigma^i\bsigma^j - \delta^{ij}/N$ and
thus it is natural to consider in this case the
tensor susceptibility
\be
\chi_T \, =\, \sum_x\, \< \bsigma^i\bsigma^j(0)\, ;\, \bsigma^i\bsigma^j(x)\>
\ee
To compute the expected behaviour of $\chi_T$ in the scaling limit
$\beta\to\infty$ we must compute the anomalous dimension of the
composite spin-two operator. In the following we will generalize the
discussion and we will discuss all non-derivative dimension-zero
operators.
A suitable basis is given by
\be
{\cal O}^{(n)}_{j_1\ldots j_n}\, =\, \bsigma^{j_1}\ldots
\bsigma^{j_n}\, -\, \hbox{\rm traces}
\ee
where ``traces" must be such that ${\cal O}^{(n)}_{j_1\ldots j_n}$
is completely
symmetric and traceless. These polynomials are irreducible $O(N)$-tensors
of rank $n$ and as such they renormalize multiplicatively with
no off-diagonal mixing.
We are now going to compute the anomalous dimension of these tensors.

As before we will take advantage of the fact that this quantity has already
been
computed in the continuum up to four loops~\cite{Wegner_1,Wegner_2}. If
we indicate with $\gamma^{(n),\msbar} (\beta)$ the anomalous dimension
of ${\cal O}^{(n)}_{j_1\ldots j_n}$ and expand
\be
\gamma^{(n),\msbar} (\beta)\, =\, {\gamma^{(n)}_0\over\beta}\, +\,
{\gamma^{(n),\msbar}_1\over\beta^2}\, +\,
{\gamma^{(n),\msbar}_2\over\beta^3}\, +\,
{\gamma^{(n),\msbar}_3\over\beta^3}\, +\, O(1/\beta^4)
\ee
we have~\cite{BLZ_2,Wegner_1,Wegner_2}
\begin{eqnarray}
\gamma^{(n)}_0 &=& {n\over 4\pi} (N+n-2) \\
\gamma^{(n),\msbar}_1 &=& 0 \\
\gamma^{(n),\msbar}_2 &=& {3n\over 64\pi^3} (N+n-2) (N-2)   \\
\gamma^{(n),\msbar}_3 &=& {n\over 384\pi^4} (N-2)(N+n-2)\left[4(5-N) +
3\zeta(3)
(2-n(N+n-2))\right]
\end{eqnarray}
Of course $2 \gamma^{(1),\msbar} (\beta) =  \gamma^{\msbar} (\beta)$ as
${\cal O}^{(1)}_j=\bsigma^j$.
To compute $\gamma^{(n),latt}(\beta)$ we consider the insertion of
${\cal O}^{(n)}_{j_1\ldots j_n}$ in the two-point function at zero
external momenta. If we indicate the one-particle irreducible
correlation on the lattice with $\Gamma^{(2)}_{(n),latt}(\beta,h;1/a)$
and its counterpart in the continuum with
$\Gamma^{(2)}_{(n),\msbar}(\beta,h;\mu)$, the general results of
\cite{Brezin-LeGuillou-ZinnJustin} imply that there exists a
finite constant $Z^{(n)}(\beta,\mu a)$ such that
\be
\Gamma^{(2)}_{(n),latt}(\beta,h;1/a)\, =\,
Z_2 (Z^{(n)})^{-1}\, \Gamma^{(2)}_{(n),\msbar}(Z_1^{-1}\beta,
Z_1 Z_2^{-1/2} h;\mu)
\ee
Then using the renormalization-group equation satisfied by the
correlation function on the lattice
\begin{eqnarray}
\left[-a{\partial\over\partial a} \right. &+&\, W^{latt}(\beta)
{\partial\over\partial\beta^{-1}} +\, \gamma^{(n),latt}(\beta)-\,
\gamma^{latt}(\beta)  \nonumber \\
&+& \left. \left({1\over2}\gamma^{latt}(\beta)+
\,\beta W^{latt}(\beta)\right)\,
h{\partial\over\partial h}\right]\, \Gamma^{(2)}_{(n),latt}\, =\, 0
\end {eqnarray}
and the analogous equation valid in the continuum we get
\be
\gamma^{(n),\msbar}(Z_1^{-1}\beta)\, =\,
\gamma^{(n),latt}(\beta)\, - W^{latt}(\beta){1\over Z^{(n)}}
{\partial Z^{(n)}\over \partial\beta^{-1}}
\ee
We will now compute $Z^{(n)}(\beta)$ up to two loops
and, using this equation, we will get the first three terms in the expansion of
$\gamma^{(n),latt}$.

To perform the computation we can arbitrarily choose the indices $j_1,\ldots
j_n$. We will choose $j_1=\ldots=j_n=1$ where $1$ is the direction of the
magnetic field. We must then express
${\cal O}^{(n)}_{1\ldots 1}$ in terms of the field $\bpi$.
To obtain this expansion
let us firstly notice that ${\cal O}^{(n)}_{j_1\ldots j_n}$ satisfies
a recursion relation of the form
\be
{\cal O}^{(n)}_{j_1\ldots j_n}\, =\, {1\over n}\left(\bsigma_{j_1}\,
{\cal O}^{(n-1)}_{j_2 \ldots j_n}\, + \, \hbox{\rm perm}\right)\, +\,
\alpha^{(n)}\left(\delta_{j_1j_2}
{\cal O}^{(n-2)}_{j_3\ldots j_n}\, + \, \hbox{\rm perm}\right)
\label{recursionO}
\ee
where ``perm" indicates the permutations necessary to make
the r.h.s completely symmetric.
To obtain $\alpha^{(n)}$ let us set
\be
\sum_{j_1} \bsigma_{j_1}\, {\cal O}^{(n)}_{j_1\ldots j_n}\, =\
\beta^{(n)}\, {\cal O}^{(n-1)}_{j_2 \ldots j_n}
\ee
where $\beta^{(n)}$ is a constant to be determined.
Then, by requiring
\be
\sum_{j_1} {\cal O}^{(n)}_{j_1 j_1 j_3\ldots j_n}\, =\, 0
\ee
we get
\be
\alpha^{(n)}\, =\, - 2 \, {\beta^{(n-1)}\over n(N+2n-4)}
\ee
A second relation is obtained by multiplying (\ref{recursionO}) by
$\bsigma_{j_1}, \ldots  \bsigma_{j_n}$ and summing over all indices.
We thus get
\be
\beta^{(n)}\, =\, {N+n-3\over N+2n-4}
\ee
and
\be
\alpha^{(n)}\, =\, - {2 (N+n-4)\over n (N+2n-4) (N+2n -6)}
\ee
Let us now introduce the notation
\be
P_n(\bsigma^1)\, =\, {\cal O}^{(n)}_{11\ldots 1}
\ee
{}From (\ref{recursionO}) we get
\be
P_n(x)\, =\, x P_{n-1} (x) + {n(n-1)\over 2}\, \alpha^{(n)}\, P_{n-2}(x)
\ee
This recursion relation, together with
$P_0(x)=1$ and $P_1(x)=x$ completely defines $P_n(x)$. The solution
is given by \cite{Gradshteyn}
\be
P_n(x)\, =\, {n!\over(N+2n-4)\ldots (n-2)}\, C^{N/2-1}_n (x)
\ee
where $C^m_n(x)$ are the Gegenbauer polynomials~\cite{Gradshteyn}.
In order to get a low-temperature expansion we use the relation
\be
C^{N/2-1}_n\, =\, {(N+n-3)!\over n! (N-3)!}\, F(N+n-2,-n;(N-1)/2;(1-x)/2)
\ee
We get in this way
\be
P_n(\bsigma^1)\, = A_n\, +\, B_n\, Q_n(\bpi)
\ee
where $A_n$ and $B_n$ are (irrelevant) numerical constants and
\begin{eqnarray}
Q_n(\bpi) &=& {\bpi^2\over2} -\, {(N+n)(n-2)\over 8(N+1)} \, (\bpi^2)^2
\nonumber \\
&&\quad + {(N+n)(N+n+2)(n-2)(n-4)\over 48 (N+1)(N+3)} \, (\bpi^2)^3\,
+\,  O(\bpi^8)
\end{eqnarray}
Let us now compute $\<Q_n \bpi^a \bpi^b\>^{irr}$
on the lattice and in the continuum. The relevant graphs are
reported in Figure~\ref{fig4}.
\begin{figure}
\vspace*{0cm} \hspace*{-0cm}
\begin{center}
\epsfxsize = 0.8\textwidth
\leavevmode\epsffile{3loops4.ps}  
\end{center}
\vspace*{0cm}
\caption{
  Feynman graphs for the insertion
  of ${\cal O}$ in the two-point
 function
 respectively (a) at the tree-level, (b) at one loop, (c) at two loops.
  The dot indicates
  the operator insertion while the cross represents the vertices coming
  from the measure term in $H^{eff}$.}
\label{fig4}
\end{figure}
If
\be
\<Q_n \bpi^a \bpi^b\>^{irr} \, =\,  \delta^{ab}\left[1 +  {q_1\over\beta}
+\,  {q_2\over \beta^2}\, +\, O(1/\beta^3) \right]
\ee
we get
\begin{eqnarray}
q_{1}^{\msbar} &=&  {1\over 8\pi} \left(2 + (n-2)(N+n)\right)
\log {h\over \mu^2}\, -\, {N-1\over 8\pi} \\
q_{2}^{\msbar} &=& {1\over 128\pi^2}\left[-4(N-3) + 16(N+n)(n-2)\right.
 \nonumber \\
&& \qquad + \left. (N+n)(N+n+2)(n-2)(n-4)\right] \log^2 {h\over \mu^2}
 \nonumber \\
&& + {1\over32\pi^2}(N-2)\left[N-3-(n-2)(N+n)\right] \log {h\over \mu^2}
\nonumber \\
&& + {1\over 64\pi^2} (N-1)(N-3)
\end{eqnarray}
while on the lattice
\begin{eqnarray}
q_{1}^{latt} &=&  {1\over 8\pi} \left(2 + (n-2)(N+n)\right)
\log {h a^2\over 32}\, -\, {N-1\over 8\pi} \\
q_{2}^{latt} &=& {1\over 128\pi^2}\left[-4(N-3) + 16(N+n)(n-2) \right.
 \nonumber \\
&& \qquad + \left.
(N+n)(N+n+2)(n-2)(n-4)\right] \log^2 {ha^2\over 32} \nonumber \\
&& + {1\over32\pi^2}(N-2)\left[N-3-(n-2)(N+n)\right] \log {ha^2\over 32}
\nonumber \\
&& + {1\over 32\pi} (2 + (N+n)(n-2))(1 + r (N+1)) \log {ha^2\over 32}
\nonumber \\
&& + {1\over 64\pi^2} (N-1)(N-3)
\nonumber \\
&& -\, {1\over 32\pi} (N-3 - (N+n)(n-2)) (1 + r(N+1))
\end{eqnarray}

Comparing the two Green functions we can now compute $Z^{(n)}$. We get
\begin{eqnarray}
Z^{(n)} &=& 1 - {n\over 8\pi\beta} (N+n-2) \,\log{\mu^2 a^2\over 32}
\nonumber \\
 && +\, {1\over \beta^2}\left[ {n\over 128\pi^2} (N+n-2)\left
(4(N-1)+(n-2)(N+n)\right)\, \log^2{\mu^2 a^2\over 32}\right. \nonumber \\
&& \qquad
\left. -\, {n\over32\pi}(N+n-2)(1 + r(N+1) ) \log{\mu^2 a^2\over 32} \right]
\end{eqnarray}
We have checked that for $n=1$ we have $(Z^{(1)})^2 = Z_2$.

The anomalous dimension $\gamma^{(n),latt}(\beta)$ is then given by
\be
\gamma^{(n),latt}(\beta)\, =\,
{\gamma^{(n)}_0\over\beta}\, +\,
{\gamma^{(n),latt}_1\over\beta^2}\, +\,
{\gamma^{(n),latt}_2\over\beta^3}\, +\, O(1/\beta^4)
\ee
where
\begin{eqnarray}
\gamma^{(n),latt}_1 &=& {n\over 16\pi} (N+n-2)(1 + r(N+1))\\
\gamma^{(n),latt}_2 &=&
{n(N+n-2)\over4\pi}\left[ {N-2\over4\pi^2}
\left(4 \pi^2 G_1 + {1\over2} - {\pi^2\over8} \right)\, +\, {11\over96}
\right. \nonumber \\
&& \quad \left.+{13\over96}(N+1)r\, +{r^2\over16}(N+1)(2N+3) -
{s\over32} (N+1)(N+3) \right]
\end{eqnarray}
Let us now consider the spin-$n$ susceptibility
\be
\chi^{(n)}\, =\, \sum_x \< {\cal O}^{(n)}_{j_1\ldots j_n} (0)\, ;\,
{\cal O}^{(n)}_{j_1\ldots j_n} (x)\>
\label{eq4.114}
\ee
It satisfies the renormalization-group equation~\cite{Brezin_76}
\be
\left[-a{\partial\over\partial a}+\,
W^{latt}(\beta){\partial\over\partial\beta^{-1}}+\,
2 \gamma^{(n),latt}(\beta) \right]\, \chi^{(n)} \, =\, 0
\ee
whose general solution is
\begin{eqnarray}
\chi^{(n)} &=& C^{(n)}_\chi e^{2\beta/w_0}
\left( {w_0\over \beta}\right)^{2 w_1/w_0^2+2\gamma^{(n)}_0/w_0} \\
&& \exp\left[2\int_0^{1/\beta} dx\, \left({1\over W^{latt}(1/x)} +
\, {1\over w_0 x^2} -\, {w_1\over w_0^2 x}-\,
{\gamma^{(n),latt}(1/x)\over W^{latt}(1/x)}
- {\gamma^{(n)}_0\over w_0 x}\right)\right]
\nonumber \\
&=& C^{(n)}_\chi e^{4\pi\beta/(N-2)}
\left({2\pi\beta\over N-2}\right)^{-(2 + n(N+n-2))/(N-2)}\,
\left\{1 + {d^{(n)}_1\over\beta}+\, O(1/\beta^2)\right\}
\label{eq2.181}
\end{eqnarray}
The (non-universal) constant $C^{(n)}_\chi$ cannot be computed in
perturbation theory. However it can be estimated in the large-$N$ limit.
For the tensor susceptibility, using the results of Appendix \ref{secB},
we have
\be
C^{(2)}_\chi \, =\, {\pi\over32}\, e^{-\pi r}\left(1 +\, O(1/N)\right)
\ee
To compute the analytic corrections let us rewrite
\be
\chi^{(n)}\, =\, R^{(n)}\xi^2_V \,
\left({2\pi\beta\over N-2}\right)^{-n(N+n-2)/(N-2)}\,
\left\{1 + {e^{(n)}_1\over\beta}+\, {e^{(n)}_2\over\beta^2}+\, +
O(1/\beta^3)\right\}
\ee
The coefficients $e^{(n)}_1$ and $e^{(n)}_2$ are given by
\begin{eqnarray}
e^{(n)}_1 &=& {n\over 4\pi} {N+n-2\over N-2} [\pi-2+\pi r(N+1)]\\
e^{(n)}_2 &=& {n^2\over 32\pi^2} {(N+n-2)^2\over (N-2)^2} \left(\pi-2+\, \pi r
(N+1)\right)^2 \nonumber \\
&& + {n(n+N-2)\over N-2}\left[ (N-2) G_1 - {1\over 8\pi} - {1\over96}(3N-14)
\right. \nonumber \\
&& +\, r(N+1) \left({7\over96} - {1\over8\pi}\right)\, +\, {r^2\over 32}
(N+1)(3N+5) \nonumber \\
&& \left. -\, {s\over 32} (N+1)(N+3) \right]
\end{eqnarray}
Of course $e^{(1)}_1 = c_1$ and $e^{(1)}_2 = c_2$.
Using then the expansion of the correlation length it is trivial to
compute $d^{(n)}_1 = 2 a_1 + e^{(n)}_1$ and the contributions proportional
to $r^2$, $r^3$ and $r^4$ in $d^{(n)}_2 = 2 a_2 + a_1^2 + e_2^{(n)}
+ 2 a_1 e_1^{(n)}$.
The discussion on the reliability of the series we have presented for $\chi_V$
applies also to the general
susceptibility $\chi^{(n)}$: indeed it is easy to see that as soon as
$r$ becomes different from zero the correction term becomes extremely
large and thus the series unreliable. For instance for $N=3$, $n=2$
(i.e. for $\chi_T$) we have
\begin{eqnarray}
d_1^{(2)}  &=&  0.362 +\, 7.476\ r -\, 15.708\, r^2 + \, 9.425\, s\\
d_2^{(2)}  &=&  36.10\, r^2\, -\, 159.72\, r^3 +\, 123.37\, r^4 +\, O(r,1)
\end{eqnarray}
which shows that one cannot hope to see asymptotic scaling in $RP^2$ model
($r=1$, $s=0$) as the correction is of order 1.

\subsection{Expansion in terms of improved variables}

Let us now discuss a second approach which follows the ideas
proposed many years ago by Parisi~\cite{Parisi_Madison}.
Let us consider a local observable
${\cal O}$ which has a perturbative expansion of the form
\be
\<{\cal O}\>\, =\, \sum_{n=0}^\infty {\cal O}_n {1\over \beta^n} \label{Obeta}
\ee
and let us suppose that this mean value has been computed up to $l$-loops.
In this case one can introduce a new quantity $x = \<{\cal O}\>\, - {\cal O}_0$
which goes to zero for large $\beta$ and, assuming that ${\cal O}_1\not=0$,
one can invert (\ref{Obeta}) to get
\be
\beta =\, {1\over x} \, \sum_{n=0}^{l-1} \overline{b}_n x^n
\ee
Then, starting from the $l$-loop expression (\ref{eq2.90}) for the correlation
length and substituting
the previous formula we get
\be
\xi_V^{(exp)} = C_{\xi_V} e^{2\pi \overline{b}_1/(N-2)}
\left({(N-2)x\over 2\pi \overline{b}_0}\right)^{1/(N-2)}
\exp\left( {2\pi \overline{b}_0\over (N-2) x }\right)
\left(1 + \sum_{n=1}^{l-2} l_n x^n +
O(x^{l-1}) \right)
  \label{eq2.105}
\ee
By expressing the analytic part of the correlation length in terms of $x$ one
changes the neglected contributions and one can hope that by a proper
choice of ${\cal O}$ the terms $O(x^{l-1})$ in \reff{eq2.105} are smaller
than those $O(1/\beta^{l-1})$ in \reff{eq2.90}.
In such a case one expects to obtain a better agreement between numerical and
theoretical values of $\xi$. Of course the main problem in this strategy is the
choice
of ${\cal O}$ since this requires at least qualitative
information on the perturbative coefficients to all loops.
However, in our case we can take advantage of the fact that we can solve
the theory in
the large-$N$ limit for the particular model \reff{eq1.1},
and choose ${\cal O}$ so that to obtain a good behaviour for
large $N$. Since the large corrections in \reff{eq2.90} are connected with
those terms
which survive in the large-$N$ limit, we expect that this choice will work
reasonably well also for finite $N$.

Looking at the formulas for $N\to\infty$ (and $s=0$)
reported in Appendix \ref{secB}, we see that a good choice is
the vector energy. Indeed if $x_V = 1 - E_V$ we get
\be
\xi_V^{(exp)}\, =\, {1\over 4\sqrt{2}}\, \exp\left( {\pi\over 2 x_V}\right)
  \left[ 1 + O(e^{-\pi/4x_V}) \right]  \;.
\ee
Thus, in this case the analytic corrections (in powers of $x_V$)
are totally absent, which is the
optimal situation. Of course, for $RP^{N-1}$ models this choice cannot be made,
since $E_V = 0$. However, in this case it is enough to consider the square root
of the tensor energy. Indeed, setting $x_T = 1 - \sqrt{E_T}$, we have
\be
\xi_T^{(exp)}\, =\, {1\over 8\sqrt{2}}\, \exp\left( {\pi\over 2 x_T}\right)
  \left[ 1 + O(e^{-\pi/4x_T}) \right]  \;,
\ee
again with no analytic corrections.

Let us now give the corresponding formulas for finite $N$. In terms of $x_V$
we have
\begin{eqnarray}
\xi^{(exp)}_V
&=& \left( {e\over 8} \right)^{1/(N-2)}
\Gamma\left(1+{1\over N-2}\right) 2^{-5/ 2}
\exp\left[- {\pi\over 4(N-2)}\right]
\nonumber \\
&& \left({2(N-2)x_V\over \pi(N-1)}\right)^{1/(N-2)}
\exp\left( {\pi(N-1)\over 2(N-2)x_V}\right) \nonumber \\
&& \left\{ 1 + {2 x_V\over\pi(N-2)(N-1)}\left[{\pi\over4} - {5\pi^2\over48}
+{\pi^2\over3}J \right.\right. \nonumber \\
&& +(2\pi)^2(N-2)\left({1\over8\pi^2} + {1\over32} - G_1 + {K\over4}\right)
\nonumber \\
&& \left.\left. + (2\pi)^2 (N+1)\left({r\over24}(6J-1)\, +\,
{r^2\over16} (8J-1)\right)\right]\right\}
\end{eqnarray}
while in terms of $x_T$ we have
\begin{eqnarray}
\xi^{(exp)}_V
&=& \left( {e\over 8} \right)^{1/(N-2)}
\Gamma\left(1+{1\over N-2}\right) 2^{-5/2}
\exp\left[-{3\pi\over 4(N-2)}\right]
\nonumber \\
&& \left({2(N-2)x_T\over \pi(N-1)}\right)^{1/(N-2)}
\exp\left( {\pi(N-1)\over 2(N-2)x_T}\right) \nonumber \\
&& \left\{ 1 + {2 x_T\over\pi(N-2)(N-1)}\left[{3\pi\over4} - {17\pi^2\over48}
-{2\pi^2\over3}J \right. \right.\nonumber \\
&& +(2\pi)^2(N-2)\left({1\over8\pi^2} - {1\over96} - G_1 + {K\over4}
-{J\over12}\right) \nonumber \\
&& \left.\left.
- (2\pi)^2 (N+1)\left({r\over24}\, -\,
{r^2\over16} (8J-1)\right)\right]\right\}
\end{eqnarray}
Notice that no term proportional to $s$ appears in this parametrization.

Numerically for $N=3$ the analytic corrections in the two expansions are
\be
1 + (0.339 - 0.378\, r + 0.292\, r^2) x_V + O(x_V^2)
\ee
and
\be
1 + (-1.043 - 2.094\, r + 0.292\, r^2) x_T + O(x_T^2)
\ee
while for $N=4$ we get
\be
1 + (0.204 - 0.157\, r + 0.122\, r^2) x_V + O(x_V^2)
\ee
and
\be
1 + (-0.479 - 0.873\, r + 0.122\, r^2)x_T + O(x_T^2)
\ee
{}From these formulas one sees that with this parametrization
the corrections are much
smaller. For $0.6\ltapprox E_{V,T} \ltapprox 0.8$ we have
$0.1\ltapprox x_T \ltapprox 0.2$ and $0.2\ltapprox x_V \ltapprox 0.4$
respectively. Thus, by parametrizing the correlation length
in terms of $x_V$ we get corrections of order $5-10$ \% for $N=3$
and of order $3-7$\% for $N=4$ for all values of $r$ with
$0\leq r\leq 1$. For the variable $x_T$ the situation is
somewhat worse. Indeed in this case we have for $N=3$ (resp. $N=4$) a
correction
of order 20\% (resp. 10\%) on the $N$-vector axis ($r=0$) which
makes this expansion less reliable than the original one where the first
correction is of order 5\% (resp. 3\%). On the $RP^{N-1}$ axis however,
the expansion
in terms of $x_T$ --- which is the only one available here as
$x_V$ = 1 identically --- is better than the original one. Indeed in this
case the correction term is of order 30-50\% for $N=3$, 10-20\% for $N=4$
and goes rapidly to zero as $N$ goes to infinity.

It is also convenient to reexpress the series
for the susceptibilities
in terms of $x_V$ or $x_T$ as for large $N$ we have
\be
\chi^{(n)} \, =\, K_V^{(n)} {x_V}^n\, \exp\left({\pi\over x_V}\right)\, =\,
K_V^{(n)} {x_T}^n\, \exp\left({\pi\over x_T}\right)
\ee
with no analytic corrections. For finite $N$ we can write
\begin{eqnarray}
\chi^{(n)} &=& R^{(n)} \xi^2_V
\left({2(N-2)x_V\over \pi(N-1)}\right)^{n(N+n-2)/(N-2)} \!\!\!
[ 1 + v_1^{(n)} x_V + v_2^{(n)} x_V^2 + O(x_V^3)]             \\
 &=& R^{(n)} \xi^2_V
\left({2(N-2)x_T\over \pi(N-1)}\right)^{n(N+n-2)/(N-2)} \!\!\!
[ 1 + u_1^{(n)} x_T + u_2^{(n)} x_T^2 + O(x_T^3)]
\end{eqnarray}
The constants $u_1^{(n)}$, $u_2^{(n)}$, $v_1^{(n)}$ and $v_2^{(n)}$
are given by
\begin{eqnarray}
v_1^{(n)} &=& {4\over N-1} {n(N+n-2)\over N-2}
\left( {1\over 8} - {1\over 2\pi}\right)\\
v_2^{(n)} &=& {16\over (N-1)^2} {n(N+n-2)\over N-2}\left[(N-2) G_1 -\,
{N-2\over4}
K -\, {J\over12} -\, {1\over 16\pi} - {12 N-37\over 384} \right. \nonumber \\
&&\qquad \left. +\, {r\over 24} (N+1)(1 - 6J) +\, {r^2\over 16} (1-8J)(N+1)
\right] \nonumber \\
&& + {8\over (N-1)^2} {n^2 (N+n-2)^2\over (N-2)^2}
\left( {1\over 8} - {1\over 2\pi}\right)^2 \\
u_1^{(n)} &=& {4\over N-1} {n(N+n-2)\over N-2}
\left( {3\over 8} - {1\over 2\pi}\right)\\
u_2^{(n)} &=& {16\over (N-1)^2} {n(N+n-2)\over N-2}\left[(N-2) G_1 -\,
{N-2\over4}
K +\, {N\over12}J -\, {3\over 16\pi} + {4 N+53\over 384} \right. \nonumber \\
&&\qquad \left. +\, {r\over 24} (N+1)
                +\, {r^2\over 16} (1-8J)(N+1)
\right] \nonumber \\
&& + {8\over (N-1)^2} {n^2 (N+n-2)^2\over (N-2)^2}
\left( {3\over 8} - {1\over 2\pi}\right)^2
\end{eqnarray}
Notice that no term proportional to $s$ appears in this parametrization.
Using the explicit formulae for $\xi$ it is immediate to obtain
the complete expansion for $\chi^{(n)}$. For $N=3$ we get for the
analytic correction to $\chi_V$
\begin{eqnarray}
1 &+& x_V\, (0.541 - 0.755\,r + \, 0.584\, r^2) \\
1 &+& x_T\, (-1.223 - 4.189\,r + \, 0.584\, r^2)
\end{eqnarray}
while for $\chi_T$ we have
\begin{eqnarray}
1 + x_V\, (0.268 - 0.755\,r + \, 0.584\, r^2) \\
1 + x_T\, (0.505 - 4.189\,r + \, 0.584\, r^2)
\end{eqnarray}
In the expansion in terms of $x_V$ the corrections are quite small
(20\% at most) and thus one expects good agreement for all values of $r$.
The expansion in terms of $x_T$ is worse and
for $r\sim 1$ the correction remains quite large (30-60\% in the region
where $\xi=50-100$). However the situation rapidly improves
when increasing $N$ (for $N=4$ the correction is now of 15--25\%)
as all the analytic terms go to zero for $N\to\infty$.

\section*{Acknowledgments}
We thank Alan Sokal for many useful comments and suggestions.

\appendix
\section{Technical Details}   \label{secA.3}

We report here a few details concerning the calculation.
In the two-loop computation
there appear three relevant lattice integrals:
\begin{eqnarray}
A^{(1)}(h) &=&
\int {{\rm d}^2 p \over (2\pi)^2} {{\rm d}^2 q \over (2\pi)^2}
{\sin^2 p_\mu \over
(\hat{p}^2 + h) (\hat{q}^2 + h) (\widehat{(p+q)}^2 + h)} \\
A^{(2)}(h) &=&
\int {{\rm d}^2 p \over (2\pi)^2} {{\rm d}^2 q \over (2\pi)^2}
{\sin p_\mu \sin q_\mu \over
(\hat{p}^2 + h) (\hat{q}^2 + h) (\widehat{(p+q)}^2 + h)} \\
A^{(3)}(h) &=&
\int {{\rm d}^2 p \over (2\pi)^2} {{\rm d}^2 q \over (2\pi)^2}
{\hat{p}^2_\mu \hat{q}^2_\mu  \widehat{(p+q)}^2_\mu \over
(\hat{p}^2 + h) (\hat{q}^2 + h) (\widehat{(p+q)}^2 + h)}
\end{eqnarray}
[here $\mu$ is {\em not}\/ summed over].
The first one is easily expressed in terms of $G_1$ and $R$ as
\be
A^{(1)}(h) \, =\, {1\over2}(I(h)^2 - R + G_1) - {1\over4} I(h) \left( 1 -
{1\over \pi}
\right) +\, O(h\log^2 h)   \;.
\ee
The last one is easily done, noticing that
\begin{eqnarray}
\sum_\mu \hat{p}^2_\mu \hat{q}^2_\mu \hat{r}^2_\mu &=&
\hat{p}^2_1 \hat{q}^2_1 \hat{r}^2_1 +
(\hat{p}^2 - \hat{p}^2_1)(\hat{q}^2 - \hat{q}^2_1)
(\hat{r}^2 - \hat{r}^2_1)  \nonumber \\
&=& \hat{p}^2 \hat{q}^2 \hat{r}^2 + ( - \hat{p}^2_1 \hat{q}^2 \hat{r}^2
+ \hat{p}^2_1 \hat{q}^2_1 \hat{r}^2 + \hbox{\rm  2 permutations}) \;,
\end{eqnarray}
from which it follows that
\be
A^{(3)}(h) = {1\over 8} +\, O(h\log^2 h)  \;.
\ee
Finally, $A^{(2)}$ can easily be reduced to the preceding ones
with the help of the trigonometric identity
\be
\sin \alpha + \sin \beta + \sin \gamma\, =\, - 4 \sin {\alpha\over2}
\sin{\beta\over2}\sin{\gamma\over2}
\ee
valid for $\alpha+\beta+\gamma=0$.
We thus get
\be
A^{(2)}(h) \, =\,
 -{1\over4}(I(h)^2 - R + G_1) + {1\over8} I(h) \left( 1 - {1\over \pi}
\right) + {1\over 192} + \, O(h\log^2 h)   \;.
\ee
In particular, the integral $G_2$ of Falcioni and Treves
\cite{Falcioni-Treves},
defined in \reff{eq2.87} above, is
\be
   G_2 \;=\;  \lim_{h\to 0}  \left[ 2 A^{(1)}(h) + 4 A^{(2)}(h) \right]
       \;=\;  {1 \over 48}  \;.
\ee

In the three-loop computation of local quantities
all integrals can be expressed in terms of
\begin{eqnarray}
B^{(1)}(h) &=& \int {{\rm d}^d p {\rm d}^d q {\rm d}^d r {\rm d}^d s\over
(2\pi)^{4d}}
(2\pi)^d \delta(p+q+r+s) { \left[ \widehat{(p+q)}^2 + h\right]^2
\over  (\hat{p}^2+h) (\hat{q}^2+h) (\hat{r}^2+h) (\hat{s}^2+h)} \\
B^{(2)}(h) &=& \int {{\rm d}^d p {\rm d}^d q {\rm d}^d r {\rm d}^d s\over
(2\pi)^{4d}}
(2\pi)^d \delta(p+q+r+s) { \left[ \widehat{(p+q)}^2 + h\right]
\left[ \widehat{(p+r)}^2 + h\right]
\over  (\hat{p}^2+h) (\hat{q}^2+h) (\hat{r}^2+h) (\hat{s}^2+h)}
\end{eqnarray}
They can be expressed in terms of $J$ and $K$ as
\begin{eqnarray}
B^{(1)}(h) &=& 4 I(h)^2 - {2\over d} I(h) + K +\, O(h)\\
B^{(2)}(h) &=& 2 I(h)^2 - {K\over2} - {1\over 6d^2} + {J\over 6} +\, O(h)
\end{eqnarray}
A useful identity in the derivation is
\be
\widehat{(p+q)}^2 + \widehat{(p+r)}^2 + \widehat{(p+s)}^2\, =\,
\hat{p}^2 + \hat{q}^2 + \hat{r}^2 + \hat{s}^2 - \sum_\mu
\hat{p}_\mu \hat{q}_\mu \hat{r}_\mu \hat{s}_\mu
\ee
valid when $p+q+r+s=0$.

Finally in the computation of the four-loop $\beta$-function an additional new
integral must be evaluated
\begin{eqnarray}
C &=& \int {{\rm d}^2 p {\rm d}^2 q {\rm d}^2 r {\rm d}^2 s\over (2\pi)
^{8}}
(2\pi)^2 \delta(p+q+r+s) {1\over \hat{p}^2  \hat{q}^2 \hat{r}^2 \hat{s}^2
\widehat{(p+q)}^2} \nonumber \\
&&\qquad \sum_{\mu\nu} \hat{p}_\mu \hat{p}_\nu \hat{q}_\mu \hat{q}_\nu
\hat{r}_\mu \hat{r}_\nu \hat{s}_\mu \hat{s}_\nu \widehat{(p+q)}^2_\mu
\end{eqnarray}
It can be simplified by noticing that
\begin{eqnarray}
&& \sum_{\mu\nu} \hat{p}_\mu \hat{p}_\nu \hat{q}_\mu \hat{q}_\nu
\hat{r}_\mu \hat{r}_\nu \hat{s}_\mu \hat{s}_\nu \widehat{(p+q)}^2_\mu
\, =\, \nonumber \\
&&\qquad \sum_\mu \hat{p}_\mu^2 \hat{q}_\mu^2
\hat{r}_\mu^2 \hat{s}_\mu^2 \widehat{(p+q)}^2_\mu
\,+\, {1\over2} \widehat{(p+q)}^2\left(\sum_\mu
\hat{p}_\mu \hat{q}_\mu \hat{r}_\mu \hat{s}_\mu\right)^2 \nonumber \\
&&\qquad - {1\over2} \widehat{(p+q)}^2 \sum_\mu
\hat{p}_\mu^2 \hat{q}_\mu^2 \hat{r}_\mu^2 \hat{s}_\mu^2
\end{eqnarray}
and that
\begin{eqnarray}
&& \sum_\mu \hat{p}_\mu^2 \hat{q}_\mu^2
\hat{r}_\mu^2 \hat{s}_\mu^2 \widehat{(p+q)}^2_\mu \,=\,
\hat{p}_1^2 \hat{q}_1^2
\hat{r}_1^2 \hat{s}_1^2 \widehat{(p+q)}^2_1 \nonumber \\
&& \qquad +\, (\hat{p}^2 - \hat{p}^2_1)  (\hat{q}^2 - \hat{q}^2_1)
(\hat{r}^2 - \hat{r}^2_1) (\hat{s}^2 - \hat{s}^2_1)
(\widehat{(p+q)}^2 - \widehat{(p+q)}^2_1)
\end{eqnarray}
In this way we get
\be
C\, =\, {J\over2}
\ee

\section{Large-N limit}   \label{secB}

In this Appendix we discuss the $N\to\infty$ limit in two dimensions
of the various quantities
computed in the paper for mixed isovector-isotensor model
\reff{eq1.1}\footnote{Let us notice that in the
literature dealing with the 1/$N$
expansion one usually defines ``$\beta$" as our $\beta/N$ as the limit
$N\to\infty$ is taken at fixed $\beta/N$.}.
As it is well known the theory is parametrized
by the mass $m_0$ of the vector particle which is related to the inverse
temperature $\beta=\beta_V+\beta_T$ by the gap-equation~\cite{Magnoli-Ravanini}
\be
{\beta\over N}\, =\, {4 I(m_0^2)^2\over 4 I(m_0^2) + r[m_0^2 I(m_0^2) - 1]}
\ee
where $r$ is again the ratio $\beta_T/(\beta_V+\beta_T)$,
and $I(m_0^2)$ is defined in \reff{eq2.13}.
In the limit
$\beta\to\infty$, $m_0$ vanishes according to
\be
m_0\, =\, \sqrt{32}\, \exp\left[ -{\pi\beta\over N}\left(1 + \sqrt{1 -
{rN\over\beta}}\right)\right] + O\left(\exp(-4\pi\beta/N)\right)
\ee
At leading order in $1/N$ it is easy to compute
\begin{eqnarray}
E_V\, & = &\, {I(m_0^2)(4 + m_0^2)-1\over 4 I(m_0^2)} \\
E_T\, & = &\, {[I(m_0^2)(4 + m_0^2)-1]^2\over 16 I(m_0^2)^2}\, =\, E^2_V
\end{eqnarray}
The susceptibilities are given by
\begin{eqnarray}
\chi_V &=& \sum_x \< \bsigma_0\cdot\bsigma_x\>\, =\, {1\over m_0^2 I(m_0^2)} \\
\chi_T &=& \sum_x \< (\bsigma_0\cdot\bsigma_x)^2\>^{conn}\, =\,
- {1\over I(m_0^2)^2} {\partial I(m_0^2)\over \partial m^2_0}
\end{eqnarray}
For the general susceptibility $\chi^{(n)}$ defined in \reff{eq4.114}
we have
\begin{eqnarray}
\chi^{(n)}\, =\, I(m_0^2)^{-n}\int \prod_{i=1}^n {dp_i\over (2\pi)^2}\,
(2\pi)^2\, \delta\left(\sum_{i=1}^n p_i\right)\,
\prod_{i=1}^n {1\over \hat{p}^2_i + m_0^2}
\end{eqnarray}

One can also compute the mass gap. One can prove that, at large distances
\cite{Magnoli-Ravanini,Muller}
\begin{eqnarray}
\<\bsigma_{(0,0)}\cdot  \sum_m \bsigma_{(n,m)}\> & = & A_1\, e^{-\mu_1 n} \\
\<\bsigma_{(0,0)}\cdot  \bsigma_{(r\cos \phi,r\sin \phi)}\> & = &
{A_2\over \sqrt{r}}\, e^{-\mu_2(\phi) r} \\
\<(\bsigma_{(0,0)}\cdot  \bsigma_{(r\cos \phi,r\sin \phi)})^2\>^{conn} & = &
{A_2^2\over r}\, e^{-2\mu_2(\phi) r}
\end{eqnarray}
where
\begin{eqnarray}
\mu_1 & = &-\log \left[1 + {m_0^2\over 2} - {m_0\over 2} \sqrt{4 +
m_0^2}\right] \\
\mu_2(\phi) &=& \cos\phi\, \hbox{\rm Arcsh}(\alpha_0\cos\phi) +
\sin\phi\, \hbox{\rm Arcsh}(\alpha_0\sin\phi)
\end{eqnarray}
with
\be
\alpha_0 = {1\over2} m_0 \sqrt{8 + m_0^2} \left[ 1 + \sqrt{1 - m_0^2(8 + m_0^2)
\left({\cos 2\phi\over 4 + m_0^2}\right)^2}\right]^{-1/2}
\ee
As usual, $\mu_1 = \mu_2(0)$.

One can also compute the second-moment correlation lengths.
We get
\begin{eqnarray}
\xi_V^{(2)} & = & {1\over m_0} \\
\xi_T^{(2)} & = &{1\over \sqrt{6}} {1\over \left[- {\partial I\over \partial
m_0^2}
\right]^{1/2}}\, \left[\left(1 + {m_0^2\over 4}\right) {\partial^2
I(m_0^2)\over
\partial (m_0^2)^2}  + {1\over2} {\partial I(m_0^2)\over \partial m_0^2}
\right]^{1/2}
\end{eqnarray}

In the perturbative limit $m_0^2\to 0$ we get
\begin{eqnarray}
E_V &= & 1 - {N\over 2\beta} \left( 1 + \sqrt{1 - {rN\over \beta}}\right)^{-1}
+ \, O(\exp(-4\pi\beta/N)) \nonumber \\
& = & 1 - {N\over 4\beta} - r{N^2\over 16\beta^2} -
r^2 {N^3\over 32\beta^3} + O(1/\beta^4)  \label{eqC.15} \\[3mm]
E_T & = & \left[1 - {N\over 2\beta} \left( 1 +
\sqrt{1 - {rN\over \beta}}\right)^{-1} \right]^2
+ \, O(\exp(-4\pi\beta/N)) \nonumber \\
& = & 1 - {N\over 2\beta} + {N^2\over 16\beta^2} - r{N^2\over 8\beta^2}
+ r {N^3\over 32\beta^3} - r^2 {N^3\over 16\beta^3} + O(1/\beta^4)
 \label{eqC.16}
\end{eqnarray}
For the susceptibilities we get in the same limit
\begin{eqnarray}
\chi_V &= &{N\over 16\beta}\, \exp\left[{2\pi\beta\over N}\left(1 +
\sqrt{1 - {r N\over \beta}}\right)\right]\, \left[1 +
\sqrt{1 - {r N\over \beta}}\right]^{-1} \nonumber \\
& = & {N\over 32\beta} e^{-\pi r} e^{4\pi\beta/N}\, \left[1 +
\left({r\over4}-{\pi\over4}r^2\right){N\over\beta}  + O(1/\beta^2) \right]
\label {eqC.17} \\
\chi_T &= &{N^2\over 32\pi\beta^2}\, \exp\left[{2\pi\beta\over N}\left(1 +
\sqrt{1 - {r N\over \beta}}\right)\right]\, \left[1 +
\sqrt{1 - {r N\over \beta}}\right]^{-2} \nonumber \\
& = & {N^2\over 128\pi\beta^2} e^{-\pi r} e^{4\pi\beta/N}\, \left[1 +
\left({r\over2}-{\pi\over4}r^2\right){N\over\beta} + O(1/\beta^2)\right]  \\
\chi^{(n)} &=& {C^{(n)}\over32} \left(2 N\over\beta\right)^n \,
\exp\left[{2\pi\beta\over N}\left(1 +
\sqrt{1 - {r N\over \beta}}\right)\right]\, \left[1 +
\sqrt{1 - {r N\over \beta}}\right]^{-n} \nonumber \\
& = & {C^{(n)}\over 32} \left(N\over\beta\right)^n
e^{-\pi r} e^{4\pi\beta/N}\, \left[1 +
\left({n r\over4}-{\pi\over4}r^2\right){N\over\beta} + O(1/\beta^2)\right]
\end{eqnarray}
where
\be
C^{(n)}\, =\, \lim_{m_0\to 0}\, m_0^2
\int \prod_{i=1}^n {dp_i\over (2\pi)^2}\,
(2\pi)^2\, \delta\left(\sum_{i=1}^n p_i\right)\,
\prod_{i=1}^n {1\over \hat{p}^2_i + m_0^2}
\ee
Explicitly $C^{(1)}=1$, $C^{(2)}=1/4\pi$, $C^{(3)}=R\approx 0.0148430$.

Let us finally consider the correlation length. It is immediately seen that
both $\mu_1$ and $\mu_2(\phi)$ converge to $m_0$ when $\beta\to\infty$. Thus
\begin{eqnarray}
\xi_V^{(exp)}\, =\, \xi_V^{(2)} \;=\; {1\over m_0} & = & {1\over4\sqrt{2}}\,
\exp\left[{\pi\beta\over N}\left(1 +
\sqrt{1 - {r N\over \beta}}\right)\right]
 \left\{ 1 + O(e^{-4\pi {\beta}/N}) \right\} \nonumber \\
& = & {1\over 4\sqrt{2}} e^{-\pi r/2} e^{2\pi\beta/N}
\left(1-{\pi\over 8} {r^2 N\over\beta} + O(1/\beta^2) \right)
  \label{eqC.19}
\end{eqnarray}
Analogously for the tensor correlations we get
\begin{eqnarray}
\xi_T^{(exp)} & = & { 1\over 2 m_0} \\
\xi_T^{(2)}   & = & { 1\over \sqrt{6} m_0} \, =\, \sqrt{{2\over3}} \,
\xi_T^{(exp)}
\end{eqnarray}

\end{document}